\documentclass[12pt,a4paper]{article}
\usepackage{epsf}
\usepackage{rotating}
\begin{document}
\title{Microscopic measurement 
of the linear compressibilities of two-dimensional fatty acid mesophases.}
\author{C\'ecile Fradin$^1$, Jean Daillant$^1$, Alan Braslau$^1$,
Daniel Luzet$^1$, Michel Alba,$^1$ \\
Michel Goldmann$^{2,3}$\\
$^1$ Service de Physique de l'Etat Condens\'{e},
CEA Saclay,\\
F-91191 Gif-sur-Yvette Cedex, France.\\
$^2$ LURE, F-91405 Orsay Cedex, France.\\
$^3$ Laboratoire Physico-Chimie Curie, Institut Curie,\\
11 rue Pierre et Marie
Curie, 75231 Paris Cedex 05.\\}
\date{\today}
\maketitle
\begin{abstract}
The linear compressibility of two-dimensonial fatty acid mesophases
has been determined by grazing incidence x-ray diffraction.
The unit cell parameters of the $L_2$, $L_2'$, $L_2''$, $S$ and $CS$ phases of
behenic acid $CH_3-(CH_2)_{20}-COOH$ and of the $L_2$ phase of myristic
acid $CH_3-(CH_2)_{12}-COOH$ were determined as a function of surface 
pressure and temperature. Surface pressure vs molecular area isotherms
were reconstructed from these measurements, and the linear compressibility
(relative distortion along a given direction for isotropic applied stress) was 
determined both in the sample plane and in a plane normal to the aliphatic
chain director (transverse plane). The linear compressibilities range 
over two orders of magnitude from 0.1 to 10 m/N and are distributed  
depending on their magnitude in 4 different sets which we are able
to associate with different molecular mechanisms.
The largest compressibilities (10 m/N) are observed in the tilted phases.
They are apparently independent on the chain length and could be related
to the reorganization of the headgroup hydrogen-bounded network, whose
role should be revalued. 
Intermediate compressibilities are observed in phases with quasi long-range 
order (directions normal to the molecular tilt in the $L_2$ or $L_2'$
phases, $S$ phase, and could be related to the ordering of these phases.
The lowest compressibilities are observed in the solid untilted $CS$ phase and for
one direction of the $S$ and $L_2''$ phases. They are similar to the 
compressibility of crystalline polymers and correspond to the interactions 
between methyl groups in the crystal. Finally, negative compressibilities
are observed in the transverse plane for the $L_2'$ and $L_2''$ phases and can be
traced to subtle reorganizations upon untilting.
\end{abstract}

\vskip 1truecm
\noindent {\it submitted to EPR B}\\

\vglue 1.7truecm
\noindent PACS:\\
\noindent {\bf 61.10-i~~~} X-ray diffraction and scattering.\\
\noindent {\bf 68.10.Et} Fluid surfaces and fluid-fluid interfaces; 
interface elasticity,
viscosity and viscoelasticity.\\
\noindent {\bf 68.60.Bs} Physical properties of thin films, nonelectronic,
mechanical and acoustical properties.\\

\noindent Short title: {\it Linear compressibilities of two-dimensional mesophases.}\\

\noindent E-mail: {\it daillant@spec.saclay.cea.fr}\\
\vfill

\section{Introduction}
The organization and phase transitions of Langmuir films,
i.e. insoluble amphiphilic monolayers at the air/water interface 
are strongly affected by the dimension 2 \cite{Renault,Berge}. 
Considerable effort has been directed during the last few years 
towards the determination of the structure of the different 
phases of those systems by grazing incidence x-ray 
diffraction (GID)\cite{Barton,Lin,Kenn,Bommarito}. 
On the theoretical side, many aspects of the phase transitions
can be understood using a recently developed Landau theory
\cite{Kag1,Kag2}, and realistic simulations are now available 
\cite{Kara,Haas}.\\

Beyond the great achievement that was the determination of the 
two-dimensional structure of Langmuir films, a next step, to 
which simulations contribute, is the understanding of the role of
intermolecular forces and conformational defects. 
Obviously, the elastic properties of monolayers crucially depend on both
potentials and defects and their understanding in terms of basic interactions
and chain conformations is a challenging goal. 
Up to now, only the two-dimensional compression and shear elastic moduli
have been the subject of detailed studies. For example, the shear modulus 
has been determined by directly applying the shear to a large 
polycrystalline sample 
\cite{Abraham}, bending a monocrystal \cite{Bercegol}, or analysing the 
shape of Bragg singularities \cite{Renault,Berge}. Because of their
inhomogeneity, the elastic moduli obtained with  large polycrystalline samples
are much smaller than those obtained with monocrystals.
This is also true for the compression modulus. This was recognized by 
Bommarito et al. \cite{Bommarito} who determined the compression modulus
in different phases of behenic acid by measuring the surface pressure using
a Wilhelmy balance and the molecular area from GID data.
Another problem shared by all these studies (with a notable exception 
\cite{Arnaud}) is that the anisotropy of the monolayer is not considered,
preventing a precise understanding in terms of microscopic 
structure.\\
In this study, we extend the method of Ref.\cite{Bommarito} to determine
the linear compressibilities of two different fatty acids, while addressing
the anisotropy of the monolayer. 
This is possible because
the GID measurements allow the determination of the two-dimensional 
lattice parameters, the monolayer thickness and the molecular tilt.

\section{Linear Compressibility of a 2D crystal}

   The surface (volume) compressibility $\chi$ of a 2D (3D) crystal  
is defined as the relative diminution of its area (volume)
$A$ when submitted to a pressure $\Pi$:  
$\chi=-1/A(\partial A/\partial \Pi)$.
A linear compressibility
can also be defined along each direction $i$ as:
$\chi_i=-1/u_i(\partial u_i/\partial \Pi)$, where $u_i$ is 
a length along the direction $i$.
In a non-isotropic medium, the linear compressibility 
depends on the direction, and can give hints about microscopic
phenomena.
This has been successfully applied 
to polymer crystals \cite{Tashiro}, where information 
about molecular interactions was obtained by comparing 
measurements and calculations 
of the Young modulus 
and of the linear compressibility
along the different crystal axes.\\
   
   The compressibility can be expressed as a function of
the coefficients of the rank 4 elastic tensor $s_{ijkl}$ relating  
the strain tensor $u_{ij}=1/2 \left(\partial u_i / \partial x_j +
\partial u_j / \partial x_i  \right)$, describing the relative
distortion of the crystal, and the stress tensor $\sigma_{ij}$
\cite{Nye} : 
\begin{eqnarray}
 u_{ij}=s_{ijkl}\sigma_{kl}
\end{eqnarray}
(summation over repeated indices is assumed). 

The volume compressibility is
$\chi=u_{ii}/\Pi$ if $\Pi$ is the applied isotropic pressure. The linear 
compressibility in the direction of the unit vector {\boldmath $\epsilon$}
is: $\chi_{\mbox{\boldmath $\epsilon$}}=u_{ij}\epsilon_i\epsilon_j/\Pi$.
The pressure $\Pi$ applied to our systems is
on average\cite{Landau} a 2D lateral
isotropic pressure, so that, if we take ${\bf x_1}$ and ${\bf x_2}$ in the
plane of the sample, the stress tensor becomes : 
\begin{eqnarray}
\sigma_{ij}=-\Pi \delta_{ij}.
\end{eqnarray}
($\delta$ is Kronecker's symbol).
Hence $u_{ij}=-\Pi(s_{ij11}+s_{ij22})$, and
\begin{eqnarray}
\chi_{\mbox{\boldmath $\epsilon$}}=(s_{ij11}+s_{ij22})\epsilon_i\epsilon_j.
\end{eqnarray}

Owing to its symmetry properties, the rank 4 elastic tensor 
$s_{ijkl}$ can be represented by a $6 \times 6$ symmetrical matrix $s_{ij}$
with the following correspondance rules: 
$11 \> \to \> 1$, $22 \> \to \> 2$, $33 \> \to \> 3$, 
$23$ and $32 \> \to \> 4$, $13$ and $31 \> \to \> 5$, $12$ and $21 \> \to \> 6$ 
\cite{Nye}. \\

The crystals under consideration here can 
alternately be viewed as very thin three-dimensional crystals if 
their properties along the vertical direction are considered or as 
two-dimensional crystals if we are only interested in their
in-plane properties. Of course, the 3D description is more
complete and rich.\\
All the studied mesophases are either monoclinic ($L_2,L'_2,L''_2$) or 
orthorombic ($S$ and $CS$) (cf Fig.\ref{orthomono}). 
Whereas the monoclinic elastic matrix has 13 independent coefficients,
the orthorombic elastic matrix has only 9 independent coefficients.
If the axes are chosen as indicated in Fig.1.b,
the resulting form of the linear compressibility for these phases is :
\begin{eqnarray}
\chi_{\mbox{\boldmath $\epsilon$}}&=&(s_{11}+s_{12})\epsilon_1^2+(s_{12}+s_{22})
\epsilon_2^2+(s_{13}+s_{23})\epsilon_3^2 \nonumber \\
&+& (s_{15}+s_{25})\epsilon_1\epsilon_3
\label{bi}
\end{eqnarray}
For orthorombic (untilted) phases the last term vanishes.
This formula allows one to predict the value of the linear compressibility
in a particular plane. If the plane of the sample is considered,
then {\boldmath $\epsilon$} can be defined by the azimuth angle $\phi$
(cf. Fig.1.a),  
$\mbox{\boldmath $\epsilon$}= \cos{\phi}\>{\bf x_1}+\sin{\phi}\>{\bf x_2}$,
and :

\begin{eqnarray}
\chi_{\mbox{\boldmath $\epsilon$}}
=(s_{11}+s_{12})\cos^2\phi+(s_{12}+s_{22})\sin^2\phi
\label{pl}
\end{eqnarray}

For monoclinic phases,
where the molecules are tilted from the vertical by an angle $\theta$, 
another interesting plane to consider is the one perpendicular
to the molecular axis (that we will call the 
transverse plane). In that plane we have
$\mbox{\boldmath $\epsilon$}=(\cos\phi/\cos{\theta})\>{\bf x_1}
+\sin{\phi}\>{\bf x_2}-\tan{\theta}\cos{\phi}\>{\bf x_3}$
(the azimuth $\phi$ is indicated Fig.1.b), and : 

\begin{eqnarray}
\chi_{\mbox{\boldmath $\epsilon$}}&=&\lefteqn{[(s_{11}+s_{12})
-(s_{15}+s_{25})\sin\theta} \nonumber \\
& &~~~~~~~~~+(s_{13}+s_{23})\sin^2\theta]/\cos^2\theta \times \cos^2\phi \nonumber \\
&+&(s_{12}+s_{22})\sin^2\phi
\label{tr}
\end{eqnarray}

This shows that if one is able to measure two linear compressibilities in the
plane of the sample, this is sufficient to yield all the linear compressibilities
in that plane, and to determine the coefficients $(s_{11}+s_{12})$ and
$(s_{12}+s_{22})$. If the molecular tilt $\theta$ can be measured 
independently, two linear compressibilities in the transverse plane 
can be evaluated,
hence all the linear compressibilities in that plane. 
One has therefore access to the coefficient 
$[(s_{11}+s_{12})-(s_{15}+s_{25})\sin\theta+(s_{13}+s_{23})\sin^2\theta]
/\cos^2\theta$. If one is  able to evaluate this coefficient in a 
large enough tilt angle range, then  
$(s_{13}+s_{23})$ and $(s_{15}+s_{25})$ can be obtained.

The two diffraction peaks generally found in GID experiments for 
orthorombic or monoclinic phases allow one 
to measure two such linear
compressibilities in the plane of the sample, and the tilt angle
can be determined from the Bragg-rod profiles. 
It is then, in principle, possible to have access to all these combinations of
coefficients.

\section{Experimental Technique}

The grazing incidence diffraction (GID) experiments were carried out at the
D41B beamline of the LURE-DCI storage ring in Orsay, France.
The $\lambda= 0.1488 nm$ 
radiation was selected using a $Ge(111)$ monochromator.
The grazing angle of incidence $\theta_i = 2.09 mrad$ was fixed slightly
below the critical angle for total external reflection using a mirror.
The width and height of
the incident beam were fixed by slits.
The diffracted radiation with in-plane wave-vector transfer $q_{xy}$ 
was selected using a Soller collimator (opening $1.43 mrad$, i.e. 
$0.0056 \AA^{-1}$ at $1.5 \AA^{-1}$),
and detected in a vertically mounted argon-filled 
position sensitive detector (PSD)\cite{Lem}.

The Langmuir trough mounted on the diffractometer was equipped
with a movable single barrier,
allowing the compression of the film. The surface pressure $\Pi$ 
was measured using a Whilhelmy balance. 
The vessel containing the trough was sealed, pumped and filled with
a flow of water saturated helium.
The temperature was regulated to within $\pm 0.5^{\circ} C$ using
the water of a large thermal bath.

The amphiphilic molecules used where fatty acids $C_{n-1}H_{2n-1}COOH$, 
purchased from Fluka ($> 99 \%$ purity), and used as obtained. 
Two different chain lengths were investigated:
behenic acid $n=22$ (we shall sometimes use $C_{22}$ for it in the following),
and myristic acid $n=14$ ($C_{14}$)
in order to get a first insight into the role of chain length (for a 
simple homogeneous solid plate the compressibility is expected 
to be proportional to the thickness).
Behenic acid was dissolved in chloroform and myristic acid in hexane 
(both from Merck, analytical grade), to a concentration of $1g/l$,
and approximately $100 \mu l$ of the solution was carefully spread  
on the water surface.\\
The film was then compressed step-by-step, 
and Bragg peaks were recorded at each (fixed) pressure step. The total time
required to record the Bragg peaks and Bragg rods was typically
3 hours per step. \\

\section{Results}

\subsection{Characterization of the unit-cell parameters}

All the crystals investigated in this study have a two-dimensional
rectangular (distorted hexagonal) unit-cell (Fig.1).
In the $L_2$ (smectic $I$ according to the liquid crystal terminology)
and $L_2''$ (smectic $I'$) phases, molecules are tilted towards one 
of their six 
nearest neighbors, whereas in the $L_2'$ (smectic $F$) phase they are
tilted towards one of
their six next-nearest neighbours. In the more ordered $S$ (smectic $U'$)
and $CS$ (smectic $U$) phases, there
is no molecular tilt. Phase diagrams of behenic and myristic acids can respectively
be found in references \cite{Kenn,Bommarito} and \cite{Bibo,Akamatsu}.\\

The peak positions measured in the horizontal plane for the two different 
fatty acids at different temperatures are given Fig.\ref{pic22} and
\ref{pic14}, where the corresponding phases are indicated.

For all these phases, only the two lowest-order diffraction peaks are observed, 
corresponding to the \{1,1\} (degenerate) and \{2,0\} (non-degenerate)
Bragg reflections (cf Fig.\ref{NN}).
The $d$ spacing of the corresponding diffracting planes can be very simply
derived from the powder-diffraction peak position $q_{xy}$ (obtained 
as well as the hwhm by fitting the peak with a Lorentzian) by: $d=2\pi/q_{xy}$.
Both cell parameters $a$
and $b$, characterizing the size of the rectangular cell in the horizontal plane, 
are hence precisely determined from the diffraction data.
The molecular area is then given by $A=ab/2$.

The tilt angle can be extracted from the rod-scans in the following way.
The maximum intensity in the Bragg rod is obtained when 
the wave-vector transfer 
${\bf q}$ is perpendicular to the molecular 
axis,
whose direction is given by the director ${\bf n}$.
From the relation ${\bf q}.{\bf n}={\bf 0}$ we get: $q_z=q_{xy} 
\tan \theta \> \cos \psi$
where $\psi$ is the azimuthal angle between the projections of
${\bf q}$ and ${\bf n}$ in 
the horizontal plane. 
Hence the \{1,1\} peak, for which $\psi$ is never equal to $\pi/2$, has a non-zero
$q_z$ component in all tilted phases ($L_2$, $L_2'$, $L_2''$). On the other hand,
the \{2,0\} peak has a non-zero $q_z$
component only in the $L_2'$ phase ($\psi=0$), since $\psi=\pi/2$ in the $L_2$ and
$L_2''$ phases. The value of $q_z$, which can be obtained  
from the rod-scans, coupled to the knowledge of the tilt direction 
(which can itself be deduced from the number of peaks having a $q_z$ component),
leads to the value of the tilt angle. 

The cell parameters and tilt angle deduced from the peaks positions
are given Fig.\ref{cell22} for $C_{22}$ and Fig.\ref{cell14} for $C_{14}$.

\subsection{Peak widths and range of the positional order}

Phases might be classed according to their cristallization
direction(s) \cite{Kaganer}. Some of them (the lower temperature phases :
$CS$ and $L_2''$) are supposed
to be 2D crystals, with long-range positional order in two directions. Other
intermediate phases ($S$, $L_2'$ and $L_{2h}$) should have long-range
order in one direction (perpendicular to the tilt direction for tilted
phases), but not in the other. Finally hexatic phases ($L_{2d}$) obtained for
higher temperature have no long-range positional order at all.

The most straightforward way to track long-range positional order is to
consider the Bragg peak widths, inversely proportional to the correlation 
length of the diffracting planes. The half width at half maximum of the
peaks are plotted in Fig.\ref{peakwi}. The situation appears
to be different for the $L_2''$ and $CS$ phases (lower inset): whereas
$CS$ has two resolution-limited peaks, supporting the fact that
it is, indeed, a 2D crystal, $L_2''$ has one resolution-limited peak, and
one very narrow but not resolution limited
peak, possibly implying that cristallization along $a$
(parallel to the molecular tilt) is not perfect. The anisotropy in 
correlation lengths of the intermediate phases (middle inset) appears 
clearly with $L_{2h}$ phase, whose two peaks have very different widths. As for
$L_2''$ phase though, the expected cristallization (along $b$ this time) is
not perfect, since the corresponding \{2,0\} peak is not quite 
resolution-limited. In the $L_2''$ and $S$ phases, the anisotropy of 
cristallization is difficult to determine, since both peaks correspond
to Bragg planes in directions with
no long-range order, and are consequently broad. The increase
in correlation length can although be noticed, when passing from $L_2'$ to
$S$. The hexatic phase ($L_{2d}$, upper inset) has inequally broad
peaks, showing that though neither principal direction possess long-range order,
one of them (perpendicular to the molecular tilt) is ordered on a longer
scale.

\subsection{Isotherms}

It is interesting to compare the isotherms obtained  by 
determining the molecular area from
the trough area and the quantity of product spread  with
those derived from the x-ray measured molecular area. 
This is presented in Fig.\ref{iso1} and Fig.\ref{iso2}.
One can see that the pressure rises
later for microscopically determined areas, owing to the fact that the film 
is inhomogenous. This point has been stressed in Ref.\cite{Bommarito};
the actual (x-ray measured)
molecular area is in fact smaller than the assumed molecular area,
which takes into account large surfaces deprived of film.
However, the molecular areas obtained using the trough 
area and the spreaded amount are smaller at higher pressure 
than those determined by using the x-ray derived molecular areas:
this is due to the loss of amphiphilic
molecules, either in the subphase, through the barriers, or by collapse.
For all these reasons, one can see that compressibilities cannot be
reliably determined 
from a standard isotherm, and that only diffraction data
can bring reliable results.
Also, the diffraction data can give access to the anisotropic 
linear compressibilities,
which is not the case when using the standard isotherms.

For behenic acid at $T=8^{\circ}C$, the trough isotherm shows
a $L_2'$ phase that has not been observed with x-rays, possibly because 
its range of existence is very small.

\subsection{Transverse cell}

A good way to look at the data, as was pointed out by Kuzmenko
et al. \cite{Kaganer}, is to consider the parameter
cells in the transverse plane, and to plot the transverse parameter
$b_T$ against the transverse parameter $a_T$, as illustrated in
Fig.\ref{banana}. This way of plotting the data allows one to put 
emphasis on the kind of molecular packing and 
will help us to better understand the negative
linear compressibilities obtained in the transverse plane (see below).

Phases are distributed on an arc, the extremities of which 
correspond to densely
packed phases (with backbone ordering), and center to less densely packed 
phases (with no backbone ordering). There are two possible backbone
arrangements for carbon chains \cite{Kaganer} : herringbone (HB)
or pseudo-herringbone (PHB) as  shown in Fig.\ref{PHBHB}.
Backbone packing strongly influences
the size and shape of the rectangular cell, which is why backbone
ordered phases are distributed on either end of the arc, depending on
whether they achieve HB or PHB packing.  Almost all phases
possessing backbone ordering (ie. in our case $L_2'$, $L_2''$, $S$
and $CS$) have HB packing. The only known phase with PHB packing is
a particular type of the so-called $L_2$ phase, $L_{2h}$ \cite{Kaganer}.
The other type $L_{2d}$ possesses no backbone ordering, and in 
consequence has a quasi hexagonal cell ($b=\sqrt3a$), which places it
in the middle of the arc. Upon decrease of temperature, the system stays
on the arc, passing from its center (undistorted unit-cell, 
molecules free to rotate around
their axis), to one of its ends (molecules with a definite azimuthal 
position).

The rescaling with chain length of the phase diagram of fatty acids 
\cite{Bibo} implies that for a same temperature, molecules with
shorter chains will organize in less dense phases (i.e. at higher $a_T$ and
$b_T$, placing them more in the center of the arc). 
This is in complete agreement with our results,
which place the $L_2$ phases of $C_{14}$ almost exactly in the middle
of the arc (making them $L_{2d}$ phases, perhaps with a faint PHB
ordering), and all the phases of $C_{22}$ towards the extremities. 
For $C_{14}$, $4^{\circ} C$ gives acces to ``high temperature phases'',
whereas for $C_{22}$, $20^{\circ} C$ still leads to ``low temperature
phases''.

The first interesting remark about Fig.\ref{banana} is that for the
same temperature, under compression, the transverse cell roughly keeps
its area (deplacement along a line where the product $a_Tb_T$ is
constant), but it jumps from one end of the arc (PHB) to the other (HB)
when passing from $L_{2h}$ to $L_2'$ or $L_2''$. These two particular phase
transitions involve a complete rearrangement of the cell around its
central molecule, since not only the backbone packing, but also the
transverse cell dimensions undergo a discontinuous change (this
discontinuity does not exist in the in-plane parameters $a$ and $b$).
This jump is of particular interest to us, since it might explain 
some unexpected features of the compressibilities we measured, as will
be explained below.

The second interesting remark about Fig.\ref{banana} is that the spread
of points of the different phases have very different extensions and directions.
This can be related to an anisotropy in transverse compressibility (that will
be developed in the next section), and compared with the anisotropy in
cristallization.
The $L_{2d}$ phases roughly extend in the direction of a ``constant geometry''
line (for which the ratio $b/a$ stays constant), showing an
equivalent compressibility in both the $a_T$ and $b_T$ direction. 
The $L_{2h}$ and $L_2''$ phases extend in a direction of constant $b_T$,
indicating a stronger compressibility along $a_T$, whereas the $L_2'$
and $S$ phases extend in a direction of constant $b_T$, related to a larger
compressibility along $a_T$. For the $CS$ phases, the extension of the spread
of points on the diagram is almost null.

\subsection{Compressibilities}

The different compressibilities calculated from our data are given in
table \ref{tab} for the two different compounds and in the different phases 
studied. 
The compressibilities range from $0.1$ to
$8m/N$, and, surprisingly, 
three distinct orders of magnitude can be clearly identified:
around $6m/N$ for the tilted phases, $0.6m/N$
for the $S$ phase, and $0.2m/N$ for the $CS$ phase.\\

Because the different phases are anisotropic, it is interesting to 
discriminate along the different directions in order to better
understand this hierarchy of compressibilities. The most interesting
linear compressibilities are those along the unit-cell axes $a$ and $b$ with
$\chi = \chi_a+\chi_b$, and the anisotropy is best visualized by building
polar diagrams representing the magnitude of the linear compressibility
along a given direction 
(i.e. the relative distortion along that direction under an isotropic
2D stress) as a function of the angle that this direction makes
with the unit-cell axes. Such polar diagrams can be 
constructed using the relations \ref{pl} and \ref{tr} of
section 2.  They are represented in Fig.\ref{ch1} 
(with the same scale) for the different phases we investigated.
Polar diagrams make it obvious that most phases 
have a highly anisotropic compressibility.
In fact, the only isotropic phase, from the 
point of view of compressibility is the $CS$ phase.
The other extreme is represented by the $L_2''$ phase for which
the linear compressibilities
$\chi_a$ and $\chi_b$ differ by two orders of magnitude.
The anisotropy might be traced 
back either to the anisotropy induced by the tilt direction (in-plane
compressibilities of the $L_2$, $L_2'$ and $L_2''$ phases), or
to the anisotropy induced by 1D cristallization like in the $S$ phase
(see also Fig.\ref{ch1b}). 
Let us recall to this point 
that tilted phases have their long-range positional order direction
perpendicular to their tilt direction (transverse $L_2$ compressibilities),
and the $S$ phase, perpendicular to the tilt direction of the
$L_2'$ phase from which it originated upon compression.\\
The central result of this paper which summarizes these data
is Fig.\ref{nuage}, where the
linear compressibilities along the $a$ and $b$ axes are 
represented on a log-log scale.
Different regions corresponding to compressibilities presumably associated 
with different mechanisms can be clearly identified on this figure, 
where the anisotropy is also most clearly visible.
The largest linear compressibilities are observed 
in the  $L_{2}$ (right, central), $L_2'$ (central, top)
and $L_2''$ (right, bottom) phases in the tilt direction.
The lowest values are observed in some phases
and directions where the layer possesses a long range positional order,
that is in both direction in the $CS$ phase (left, bottom),
in one direction of the $S$
phase (left, central), and in one direction of the $L_2''$ phase. 
They have to be related
to the compression of already well organized molecular planes.
The intermediate values are found in the transverse
plane, in one direction of the $S$ phase and of the $L_2$ and $L_2'$ phases.
They correspond to some directions
having no long-range positional order (the $b$ direction of $S$),
and others generally expected
to have it (the $b$ direction of $L_2$ and the $a$ direction of $L_2'$).\\

One can make abstraction of the effect of tilt by looking at the 
linear compressibilities in the transverse plane normal to the molecular axis.
A transverse compressibility $\chi_T=-1/A_T(\partial A_T/\partial \Pi)$,
where $A_T$ is the transverse area of the cell, and a tilt compressibility
$\chi_{\theta}=\chi-\chi_T=\tan\theta (\partial\theta/\partial\Pi$) can 
then be defined. $\chi_{\theta}$  describes 
the part of the compressibility due to the untilting of the molecules.
If the molecules are tilted along $a$, one can also 
write $\chi_a=\chi_{a_T}+\chi_{\theta}$,
or $\chi_b=\chi_{b_T}+ \chi_{\theta}$ if they are tilted along $b$.
For the orthorombic phases, the in-plane and transverse compressibilities
are of course equal ($\chi_{\theta}$=0).
The polar diagram resulting from the linear compressibilities in the 
transverse plane
are represented in Fig.\ref{ch1b}. 

Transverse linear compressibilities  
are mainly interesting  in the tilt direction where they differ
from the in-plane
compressibilities.  In the $L_2$ phase, linear compressibilities  are
reduced 
in the tilt direction from $5 m/N$ in the sample plane to $1 - 2 m/N$,
i.e. larger but on the order of the intermediate compressibilities mentioned 
above.
More unexpected is that $\chi_{bT}$ in the $L_2'$ phase
and $\chi_{aT}$ in the $L_2''$ phase are negative.\\

Finally, it is interesting to note 
the strong correspondance between linear compressibilities
and correlation lengths : The tilted phases only exhibits very short 
range positional order along the tilt direction where the compressibility
is very large, the $CS$ phase has 2D long-range order, and no noticeable
(on the diagram) compressibility in either direction, $L_{2h}$ and $S$ have 
1D cristallization, and show larger compressibilities in the direction
where they are not cristallized.

\section{Discussion}

We first summarize the theory of elasticity of thin plates as a basis for
the discussion. The four different sets of compressibilities (large
compressibilities in the tilted phases, intermediate compressibilities 
along the direction without tilt nor true  long-range order,
low compressibilities
in the crystal phases, and negative linear compressibilities 
in the transverse plane) will then be discussed with the aim of 
assigning a molecular mechanism to each of them.\\

The central result of the theory of elasticity for thin 
plates \cite{Landau} is that  the compressibility is proportional 
to the plate thickness $h$: $\chi = E h /(1-\sigma_p^2)$ where 
$E$ is the Young modulus and $\sigma_p$ the Poisson ratio. Of course this is
not necessarily a realistic model, but the same trends remain in more 
realistic approaches: If one is only interested in the interactions
between the chains 
(which up to now are believed to be the relevant part of the amphiphilic 
molecules), three essential contributions to the film free energy
must be considered\cite{Rieu}: attractive van der Waals forces  which ensure
the film cohesion, the entropy of conformation defects and their energy
(which must be included to account for phase transitions). It turns
out that all these components roughly scale proportionally to $n$
\cite{Rieu},
the number of segments, and it will therefore also be the case for
the compressibility which is the second derivative of the free energy
with respect to the area.\\

\subsection{Compressibility of the tilted phases}
The compressibilities of the tilted phases are the largest that we observe
($ > 5m/N$, see table \ref{tab}). 
The expectation that the compressibilities should scale with the 
segment number (i.e. a factor of $1.7$ between myristic and behenic acid)
is in strong contrast with our experimental observation
that the compressibilities of behenic acid 
and myristic acid in the $L_2$ phase at $8^{\circ} C$ are identical. 
Indeed all our results tend to indicate that the compressibility of 
tilted phases are remarkably independent of chain length (and also of 
film thickness since the compressibilities do not depend on the tilt angle).
This, and other experimental observations detailed below,
leads us to propose a reconsideration of  the respective roles of
aliphatic tails and headgroups
in the physics of amphiphilic films. In particular we propose here that
the compressibility of tilted phases could be due to the elasticity of the
hydrogen-bounded headgroup network. That would also explain the puzzling
observation that the compressibilities of behenic acid in the $L_2''$ 
and myristic acid in the $L_2$ phase at $5^{\circ} C$ are equal (and 
significantly different from their values at $8^{\circ} C$).\\ 

Further evidence that the role of headgroups needs to be reconsidered is 
as follows:\\
Firstly, the so-called ``universality'' of the phase diagrams is only 
relevant for molecules having 
different chain lengths but the same headgroup. As soon as different 
headgroups are considered like alcohols or esters, significant differences 
appear in the phase diagram topology \cite{Chuck}. Also striking is the 
fact that in some cases very high pressures can be achieved with molecules
that remain tilted throughout the whole phase diagram, which cannot
be understood if the chains only are relevant\cite{Graner}.\\

Secondly, the data of Fig.\ref{transv} add very interesting evidence to 
this interpretation.

Whereas at a given surface pressure, the molecular areas of behenic acid
and myristic acid are very similar, as one would expect if the interactions 
between headgroups are dominant and fix the area, this is not the case for 
the transverse area which differ by more than $1 \AA^2$:
We measured $18.75$ and $19.25 \AA^2/molecule$ at respectively $8^{\circ} C$ and
$20^{\circ}C$ for behenic acid  and $20.0$ and $20.5 \AA^2/molecule$ at 
$5^{\circ} C$ and $8^{\circ}C$ for myristic acid.
(Note that the reduction in molecular area is roughly $0.5A^2/molecule$ when 
the temperature is decreased by $12^{\circ}C$, or 
about $1.75 \AA^2/molecule$ when the chain length is increased by 8 segments, 
hence leading to the equivalence $1 \> CH_2 \equiv
5.25^{\circ}C$ in good agreement with what is obtained from the
transition temperatures in ref.\cite{Bibo}, around $5$ to $10^{\circ}C$ per 
$CH_2$ group.)
Moreover the tilt angle is larger for behenic acid than for myristic acid 
at the same pressure in the $L_2$ phase  (on the order of $20-25^{\circ}$ 
for myristic acid and of $25-35^{\circ}$ behenic acid).\\
If the molecular area is mainly to be fixed 
by the headgroup interactions, one does expect the same molecular area
whatever the chain length as is observed, and the differences in cross-section 
and tilt angle can be understood as follows:
Because the van der Waals interactions are stronger between $C_{22}$ 
chains than between $C_{14}$ chains due to their larger length, the number 
of defects is smaller and their cross-section is also smaller (i.e. the 
transverse area as experimentally observed). It is therefore necessary 
to have a larger tilt angle to achieve the projected area than for shorter, 
more disordered chains with a larger cross-section.\\

Finally, the temperature dependance of the compressibilities of the 
tilted phases presents interesting features.
Whereas the compressibility of behenic acid in the $L_2$ phase only slightly
decreases from $20^{\circ}$ to $8^{\circ}$, the compressibility
of myristic acid significantly increases between $8^{\circ}$ and $5^{\circ}$
from $5 m/N$ to $8m/N$ (the same happens for behenic acid but the phase at 
$5^{\circ}$ is now $L_2''$). The same increase is reported in 
Ref.\cite{Bommarito} for the $L_2$ and $L_2'$ phases of behenic acid.
This could be related to the anomalous properties of water near 
$4^{\circ}$ due to the hydrogen bound network. 

\subsection{Intermediate compressibilities}

The intermediate compressibilities obtained in the $L_2$, $L_2'$, 
$L_2''$ and $S$ phases along the directions where there is neither 
tilt nor long range order are on the order of $0.5 m/N$.
It has been proposed on the basis of a Landau theory of weak 
crystallization in Ref.\cite{Kag1,Kag2,Kaganer} 
that the absence of true long range order could be related to the 
imperfect ordering of the backbone planes. The compressibility 
would then be due to the ordering of the molecule positions.
An order of magnitude of the corresponding energy can be obtained from
our data by comparing the compressibilities of the $S$ phase and
$CS$ phase along $b$, since they both have the same cell geometry and
packing (HB), but the $S$ phase has true  long-range order 
only along $a$, whereas the $CS$ phase is a 2-D crystal. 
If one assumes that the energy $\delta E$ necessary to squeeze out the defects 
is equal to the elastic  energy necessary reduce the cell parameter 
$b$ from its large value in the $S$ phase to its low value in the $CS$ phase,
then $\delta E= \delta b^2 / \chi_b $, using a simple spring model.
$\delta b$ is the variation of the $b/2$ parameter of the cell,
that goes from $(7.6/2)$ to $(7.4/2)\AA$ for one molecule 
during the transition. This variation is representative of the transition from
a disordered state to a crystallized state in the b-direction.
$\chi_b$ is the linear compressibility along $b$ in the $S$ phase
$ \approx 0.5m/N$. With these numbers, one obtains
$\delta E \approx 2\times10^{-22} J$, 
to be compared to the the thermal energy $k_BT = 4.1\times10^{-21} J$,
indicating that defects are rather rare, which is consistent with a 
coherence length of $\approx 30$ interatomic distances in the ill-crystallized
direction of the $S$ phase.

\subsection{Compressibility of crystal phases}

The lowest compressibilities where obtained in the $CS$ phase 
and along the ``well crystallized'' directions $b$ in the $L_2''$ 
phase and $a$ in the $S$ phase. They are on the order or smaller than 
$0.1 m/N$.
It is interesting to compare these values to those obtained for
3D polymer crystals, in particular for orthorombic polyethylene 
which has a structure similar to that of the $CS$ phase. 
The values of the linear compressibilities for orthorombic polyethylene 
reported in Ref.\cite{Tashiro} are $1.8 \times 10^{-10}m^2/N$  along $a$
and $ 1.4 \times 10^{-10}m^2/N$ along $b$.  
Let us note that the 2D pressure $\Pi$ is 
homogenous to $N/m$ and not $N/m^2$, which will cause our elastic 
coefficients to be in $m/N$. To be compared to the elastic coefficients
of similar 3D materials, the bidimensional compressibilities 
will have to be multiplied by the thickness
of the layer (here $2.4 nm$).
We obtain $ \approx (2.4 \pm 0.3) \times 10^{-10}m^2/N$  along $a$
or $b$ in quite good agreement with the values for polyethylene. 
The fact that the compressibility of a film is slightly larger than that of
a 3D crystal can probably be explained by a larger number of defects. 
In any case, the compressibility of the 3D polyethylene crystals could
be nicely estimated by using only a pairwise intermolecular potential 
including a short range repulsive exponential potential and long range 
attractive van der Waals forces for a perfect crystal order with no 
defects, and one confidently assign the same molecular origin to the
lowest compressibilities of fatty acids, i.e. repulsive interactions 
between methyl groups in a well crystallized solid.

\subsection{Negative transverse linear compressibilities}

Negative transverse linear compressibilities are only observed in the
$L_2'$ and $L_2''$ tilted phases.
We propose an explanation for the negative linear compressibilities 
on the basis of the transition between PHB and HB packing as follows: 
Looking at Fig.\ref{pass}, one can see that the transition
from $L_2$ to $L_2'$ or $L_2''$ implies a dramatic change in the transverse 
cell parameters (this can also be seen in Fig.\ref{banana}). 
In each case one of the
transverse parameters ($b_T$ for $L_2/L_2'$ and $a_T$
for $L_2/L_2''$) increases
instead of decreasing upon compression. They happen to be exactly the same
for which the negative transverse linear compressibilities are observed 
upon further compression.
It is therefore likely that upon further compression after
the PHB to HB transition the unit cell is  
still reorganizing in the transverse plane towards a preferred
geometry, implying  negative linear compressibilities.

An important consequence is that the second mechanism
presented in Fig.\ref{pass} (along the diagonal) for the transition from
$L_2$ to $L_2'$, 
corresponding to a $90^{\circ}$ change of the tilt direction, is ruled out,
since it would involve a negative compressibility along $a_T$ in the $L_2'$
phase, and not along $b_T$. This result was already suggested by Brewster Angle
Microscopy \cite{Riviere}.

\section{Conclusion}
The linear compressibility of a two-dimensional fatty acid 
crystal has been measured for the first time. 
Surprisingly, the linear compressibilities can be nicely divided 
in 4 sets depending on their value which is characteristic of the 
phase and direction of crystallization, and a different molecular 
mechanism could be ascribed to each set. This lead us in particular 
to propose to reconsider the role of headgroups.\\
The largest compressibilities (10 m/N) are observed in the tilted phases.
They are apparently independant on the chain length and could be related
to the reorganization of the headgroup hydrogen-bounded network. 
The intermediate compressibilities observed in  
the directions normal to the molecular tilt in the $L_2$ or $L_2'$
phases and in the  $S$ phase in the direction without true long range 
order could be related to the progressive squeezing 
of defects in those phases.
The lowest compressibilities observed in the solid untilted $CS$ phase and for
one direction of the $S$ and $L_2''$ phases are similar to the 
compressibilities of crystalline polymers. They correspond to the interactions 
between methyl groups in the crystal. Finally, the negative compressibilities
observed in the transverse plane for the $L_2'$ and $L_2''$ and can be
traced back to subtle reorganizations upon untilting.\\
Work is in progress in order to solve some remaining puzzling questions:\\
Is it possible to further demonstrate that the compressibility of tilted
phases is mainly due to interactions between headgroups by measuring the 
compressibility for different headgroups and observing large variations?\\
Does compressibility really decrease with increasing temperature in those 
phases and what is the underlying mechanism?\\
What is the dependence of the compressibility on film thickness 
(i.e. chain length) in the untilted phases? \\
Answering those questions would allow a better understanding of 
the compressibility of two-dimensional Langmuir film crystals and open the 
way for a more quantitative theoretical approach.\\

{\bf Acknowledgements:} The help of C. Blot during the experiments is
gratefully acknowledged. This work greatly benefited from discussions 
with V.M. Kaganer who made us aware of the importance of the packing 
requirements for fatty acid monolayers and the manuscript benefited 
from a critical reading of P. Fontaine.\\

\noindent$^*$ Laboratoire de Physico-Chimie Curie is UMR 168 associated
to the Centre National de la Recherche Scientifique.

\clearpage
\vfill
\eject
\newpage

\begin{sidewaystable}

\begin{tabular}{|l|l|l|l|l|l|}   
      \hline
      \hline
 & \multicolumn{3}{c|}{\em Behenic acid (C22)}
 & \multicolumn{2}{c|}{\em Myristic acid (C14)} \\ \cline{2-6}
 & $20^oC$ & $8^oC$ & $5^oC$ & $8^oC$ & $5^oC$  \\ \hline
  $L_2$ & ${\bf \chi=6.02 \pm 0.21}$ &  ${\bf \chi=5.04 \pm 0.06}$ & & ${\bf
  \chi=5.05
 \pm 0.26}$ & ${\bf \chi=7.80 \pm 0.51}$ \\ 
    & $~~~\chi_a=5.45 \pm 0.28$ & $~~~\chi_a=4.55 \pm 0.13$ & &
    $~~~\chi_a=4.54 \pm 0.29$ & $~~~\chi_a=6.97 \pm 0.47$ \\
        & $~~~~~~\chi_{a_T}=1.02 \pm 0.17$       
        & $~~~~~~\chi_{a_T}=1.94 \pm 0.44$ &       
        & $~~~~~~\chi_{a_T}=1.63 \pm 0.63$       
        & $~~~~~~\chi_{a_T}=1.19 \pm 0.65$ \\
    & $~~~\chi_b=0.57 \pm 0.04$ & $~~~\chi_b=0.52 \pm 0.07$ & &
    $~~~\chi_b=0.61 \pm 0.06$ & $~~~\chi_b=0.67 \pm 0.07$ \\ \hline
  $L_2'$ & ${\bf \chi=5.04 \pm ...}$ & & & & \\
    & $~~~\chi_a=0.544 \pm ...$ & & & & \\
    & $~~~\chi_b=4.597 \pm ...$ & & & & \\ 
    & $~~~~~~\chi_{b_T}=-1.30 \pm 0.84$ & & & & \\ \hline
  $L_2''$ & & & ${\bf \chi=7.52 \pm 0.33}$ & & \\
     & & & $~~~\chi_a=7.46 \pm 0.33$ & & \\
     & & & $~~~\chi_b=0.082 \pm 0.026$ & & \\
       & & & $~~~~~~\chi_{a_T}=-0.72 \pm 0.52$ & & \\  \hline
  $S$ & ${\bf \chi=0.565 \pm 0.209}$ & & & & \\
      & $~~~\chi_a=0.114 \pm 0.178$ & & & & \\
      & $~~~\chi_b=0.496 \pm 0.261$ & & & & \\    \hline
  $CS$ & & ${\bf \chi=0.209 \pm 0.013}$ & ${\bf \chi=0.151 \pm 0.296}$ & & \\
      & & $~~~\chi_a=0.107 \pm 0.010$ & $~~~\chi_a=0.110 \pm 0.130$ & & \\
      & & $~~~\chi_b=0.103 \pm 0.006$ & $~~~\chi_b=0.092 \pm 0.120$ & & \\
      \hline
      \hline

\end{tabular}

\caption{Compressibilities (bold font), linear compressibilities
and transverse compressibilities  in the 
$L_2$, $L_2'$, $L_2''$, $S$, and $CS$ phases for behenic acid  
at $5^{\circ} C$, $8^{\circ} C$, and $20^{\circ} C$  and myristic acid at
at $5^{\circ} C$ and $8^{\circ} C$ in $m/N$.}
\label{tab}
\end{sidewaystable} 

\vfill
\eject
\clearpage
\newpage

\begin{figure}
\caption{Unit-cell geometry for an orthorombic phase (a)
and a monoclinic phase (b). The molecular tilt in (b) is towards the
nearest neighbors (NN) as in the $L_2$ or $L_2''$ phases. $\phi$ is the
azimuth angle defining the direction ${\bf \epsilon}$ in the monolayer plane
in (a) and in the transverse plane normal to the molecular tilt in (b).}
\label{orthomono}
\end{figure}

\begin{figure}
\caption{Bragg peak positions for behenic acid at three different temperatures 
$5^{\circ} C$, $8^{\circ} C$, and $20^{\circ} C$ as a function of surface
pressure. 
The filled squares indicate the position of the non-degenerate \{2,0\} peak,
while the empty circles indicate the position of the degenerate \{1,1\} peak.}
\label{pic22}
\end{figure}

\begin{figure}
\caption{Bragg peak positions for myristic  acid at two different temperatures 
$5^{\circ} C$ and  $8^{\circ} C$ as a function of surface pressure. 
The filled squares indicate the position of the non-degenerate \{2,0\} peak,
while the empty circles indicate the position of the degenerate \{1,1\} peak.}
\label{pic14}
\end{figure}

\begin{figure}
\caption{Unit-cell and kinematics of in-plane diffraction from tilted phases.
Dark grey triangles are for a molecular tilt towards nearest neighbors (NN)
and light grey triangles for a molecular tilt towards next 
nearest neighbors (NNN).
The Bragg planes are indicated as broken lines.}
\label{NN}
\end{figure}

\begin{figure}
\caption{Molecular tilt and unit-cell parameters 
for behenic acid at three different temperatures
$5^{\circ} C$ (empty grey squares), $8^{\circ} C$ (empty black circles),
and $20^{\circ} C$ (filled black squares) as a function of surface
pressure.  The
dashed lines separating the different phase regions are only a guide for the
eye.}
\label{cell22}
\end{figure}

\begin{figure}
\caption{Molecular tilt and unit-cell parameters  
for myristic acid at two different temperatures
$5^{\circ} C$ (grey squares) and $8^{\circ} C$ (black circles) 
as a function of surface pressure.}
\label{cell14}
\end{figure}

\begin{figure}
\caption{Bragg peak widths in the different phases as a function 
of surface pressure. 
The filled symbols indicate the position of the non-degenerate \{2,0\} peak,
while the empty symbols indicate the position of the degenerate \{1,1\} peak.
Top: Myristic acid at $5^{\circ} C$ (squares) and $8^{\circ} C$ (circles);
Middle: Behenic acid at $8^{\circ} C$ (squares) and $20^{\circ} C$ (circles);
Bottom: Behenic acid at $5^{\circ}$.}
\label{peakwi}
\end{figure}

\begin{figure}
\caption{Molecular area vs surface pressure isotherms for behenic acid
at three different temperatures
$5^{\circ} C$ (empty light grey diamonds and light grey line), $8^{\circ} C$ 
(empty grey circles and grey line),
and $20^{\circ} C$ (filled black squares and black line). 
The molecular area was determined from x-ray measurements (points) or
from spreaded amount and trough area (continuous lines).}
\label{iso1}
\end{figure}

\begin{figure}
\caption{Molecular area vs surface pressure isotherms for myristic acid
at two  different temperatures
$5^{\circ} C$ (empty grey diamonds and grey line) and  $8^{\circ} C$ 
(empty black circles and black line),
The molecular area was determined from x-ray 
measurements (points) or from spreaded amount and trough area (continuous
lines).}
\label{iso2}
\end{figure}

\begin{figure}
\caption{Distribution of the different phases as a function of their transverse
cell parameters: $L_2$ (empty circles), $L_2'$ (filled squares), $L_2''$ 
(inverted triangles), $S$ (empty diamonds), and $CS$ (empty squares).
The solid lines indicate a constant area and the broken lines a constant 
cell geometry.}
\label{banana}
\end{figure}

\begin{figure}
\caption{The two different types of molecular packing found in fatty acid
mesophases.
The molecular tilt is schematically suggested by representing a top and 
bottom methyl group orientated as to indicate the azimuth.
The pseudo-herrigbone (PHB) packing is found only in the $L_2$ phase. 
All other phases have herringbone (HB) packing.}
\label{PHBHB}
\end{figure}

\begin{figure}
\caption{Polar diagrams of the in-plane linear compressibilities 
in the $L_2$ phase of behenic and myristic acids, and the $L_2'$, $L_2''$, 
$S$, and $CS$ phases of behenic acid. Grey lines are used
for myristic acid in the $L_2$ phase.
The unit-cell is indicated by broken grey lines and the molecular tilt direction
by grey symbols. See also Fig.\ref{ch1b} for an enlarged plot of the linear 
compressibilities of $S$ and $CS$ phases.}
\label{ch1}
\end{figure}

\begin{figure}
\caption{Polar diagrams of the transverse linear compressibilities 
in the $L_2$ phase of behenic and myristic acids, and the $L_2'$, $L_2''$, 
$S$, and $CS$ phases of behenic acid. Grey lines are used
for myristic acid in the $L_2$ phase and the $CS$ phase of behenic acid.
The grey shaded lobes indicate a negative transverse compressibility.
The unit-cell is indicated by broken grey lines and the molecular tilt direction
by grey symbols.}
\label{ch1b}
\end{figure}

\begin{figure}
\caption{Distribution of the different phases as a function of their 
compressibilities along the two principal directions of the cell:
$L_2$ (empty circles), $L_2'$ (inverted triangle), $L_2''$ (diamond),
$S$ (empty square), $CS$ (triangles). The transverse compressibility
in the $L_2$ phase is also represented (filled black circles).
The broken line indicates isotropic compressibility ($\chi_a=\chi_b$),
whereas the dotted lines roughly separate the different regions 
of compressibility.}
\label{nuage}
\end{figure}

\begin{figure}
\caption{Molecular area (circles) and transverse molecular area 
(squares and triangles) as a function of surface
pressure. Filled symbols are for myristic acid at 
$5^{\circ} C$ and $8^{\circ} C$; empty circles for behenic acid at 
$8^{\circ} C$ (squares) and $20^{\circ} C$. Triangles are for the
$L_2$ phase and inverted triangles for the $L_2'$ phase.}
\label{transv}
\end{figure}

\begin{figure}
\caption{Transverse unit-cell geometries found in the tilted phases 
$L_2$ (top left), $L_2'$ (bottom left and right)  and $L_2''$ (top right).
The molecular tilt is schematically suggested by representing a top and 
bottom methyl group orientated as to indicate the azimuth.
The unit-cell is represented by solid lines. Note that the scale
of the $a$ and $b$ axes has been respected. Dotted lines represent 
Bragg planes with long range positional order. The $L_2$ to $L_2''$ 
phase transition and to different scenarios for the $L_2$ to 
$L_2'$ phase transitions are indicated by arrows and the change in 
unit-cell parameters is given. In the case of the $L_2$ to $L_2''$ 
phase transition, the scenario with the broken arrow can be ruled-out 
(see text for details); note that $a$ and $b$ are exchanged during 
the observed transition.}
\label{pass}
\end{figure}

\begin{figure}
\caption{Schematics of the three suggested molecular mechanisms for
(a) the compressibility of the tilted phases where the hydrogen bounded 
network of headgroups is proposed to play an important role, 
(b) the linear compressibility along directions of imperfect 1D 
crystallization where positional defects are important, and 
(c) the $CS$ phases and directions of perfect 1D crystallization where
the compressibility is due to the repulsive interaction between 
well-crystallized methyl groups.}
\label{meca}
\end{figure} 

\clearpage

\vfill
\eject

\newpage

\epsfxsize 14truecm
\hfil \epsfbox{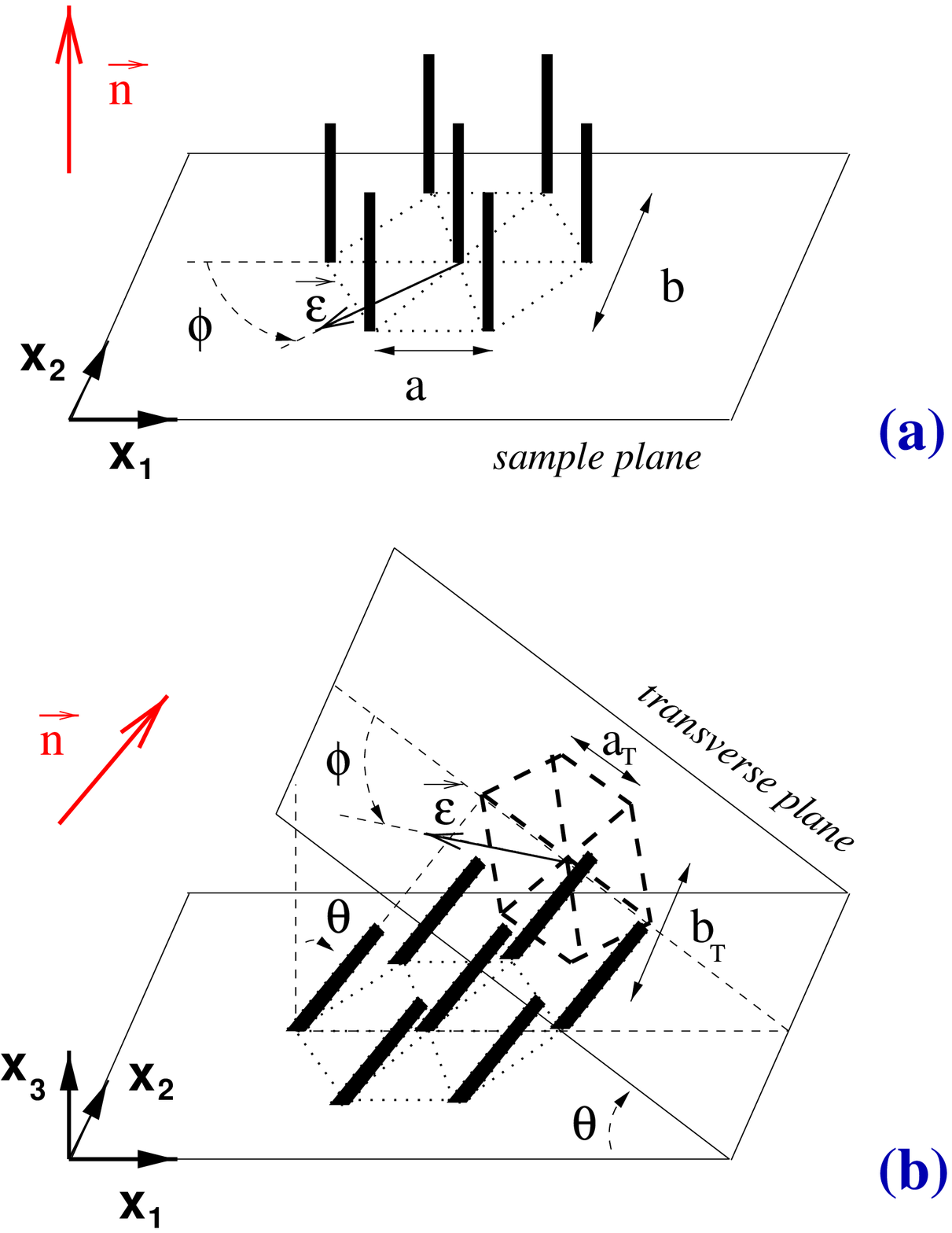} \hfil
\vskip .1truecm
\centerline{Fig. 1}
\newpage

\epsfxsize 14truecm
\hfil \epsfbox{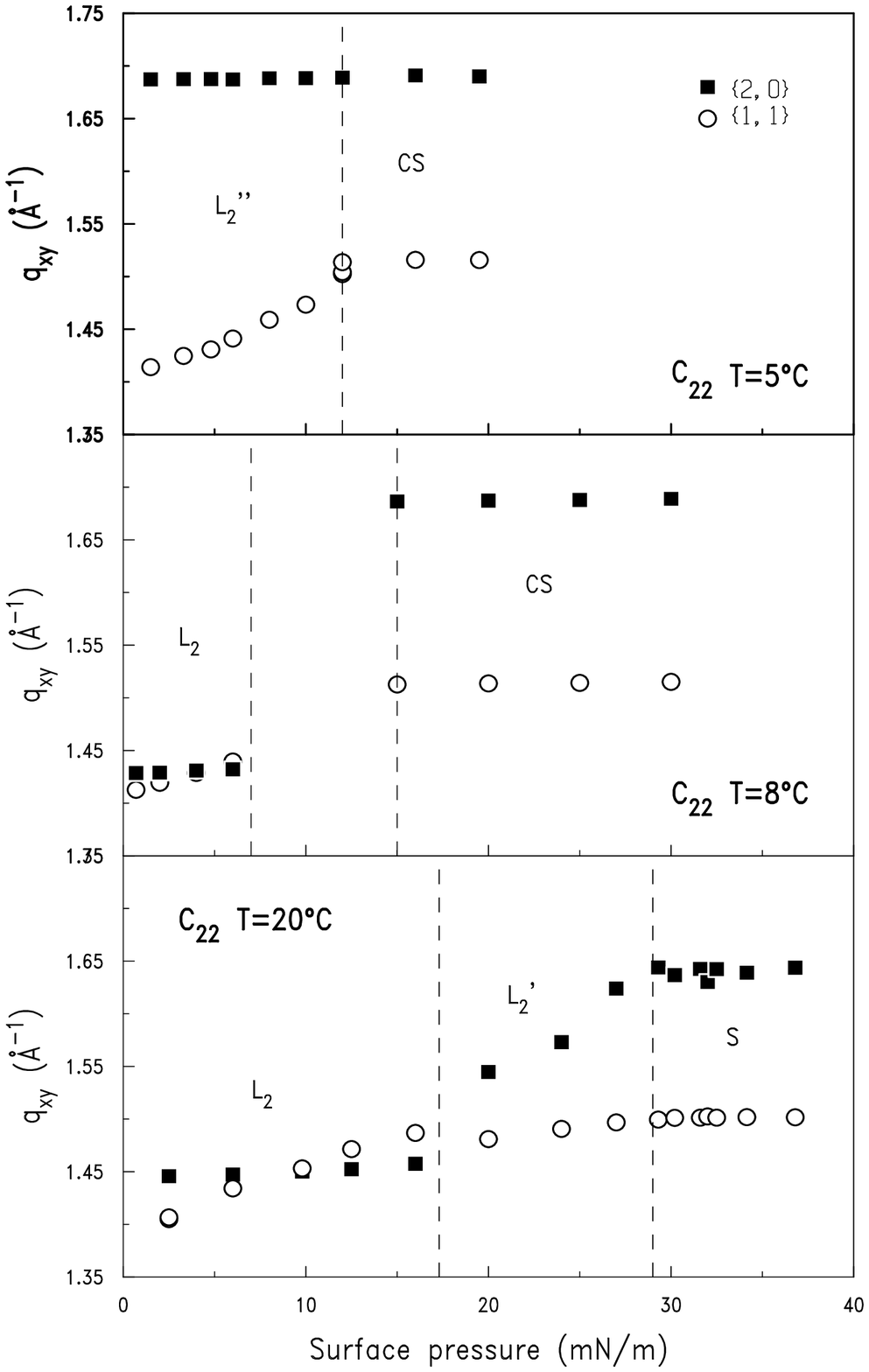} \hfil
\vskip 1truecm
\centerline{Fig. 2}
\newpage

\epsfxsize 14truecm
\hfil \epsfbox{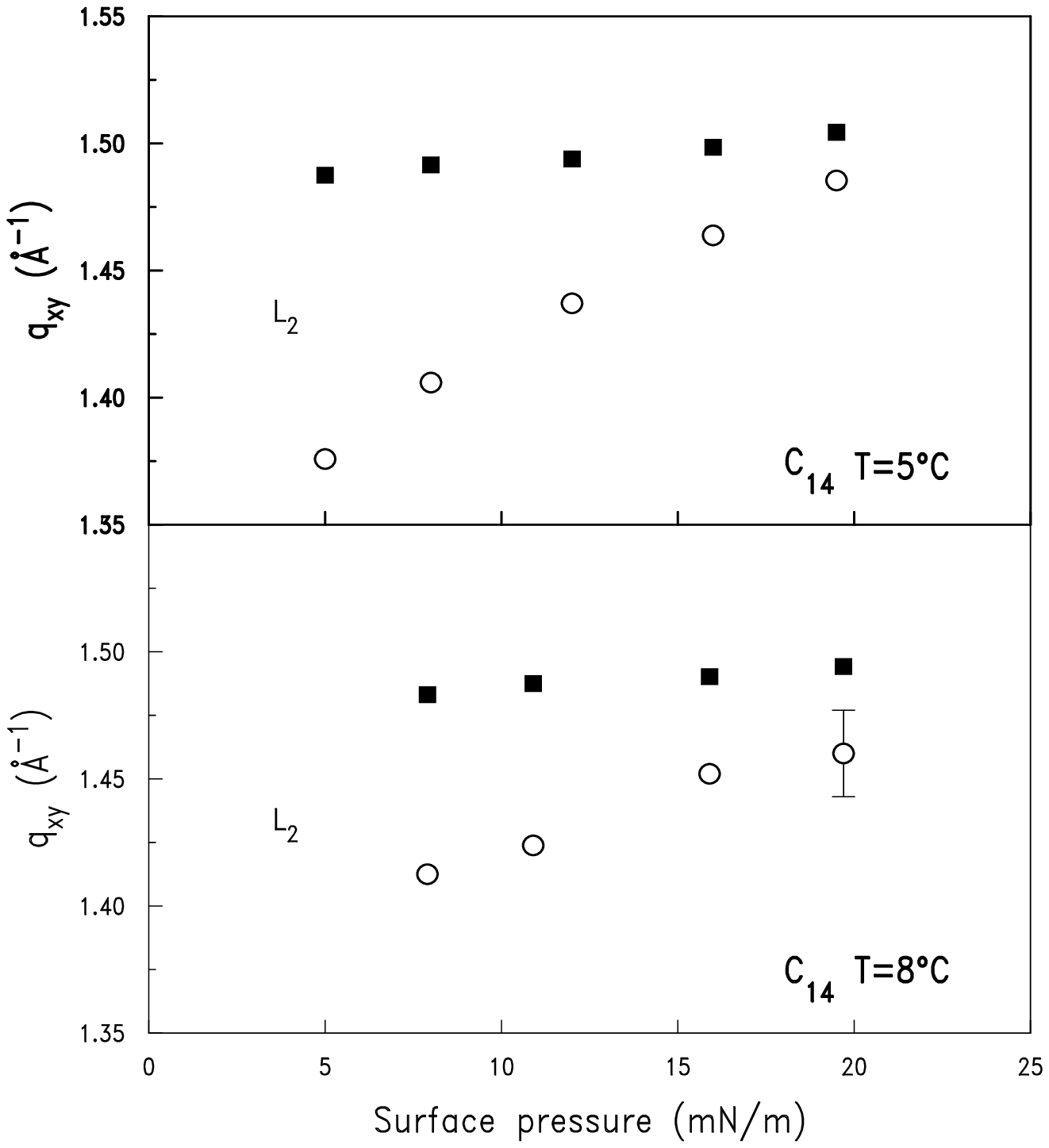} \hfil
\vskip 1truecm
\centerline{Fig. 3}
\newpage

\epsfxsize 14truecm
\hfil \epsfbox{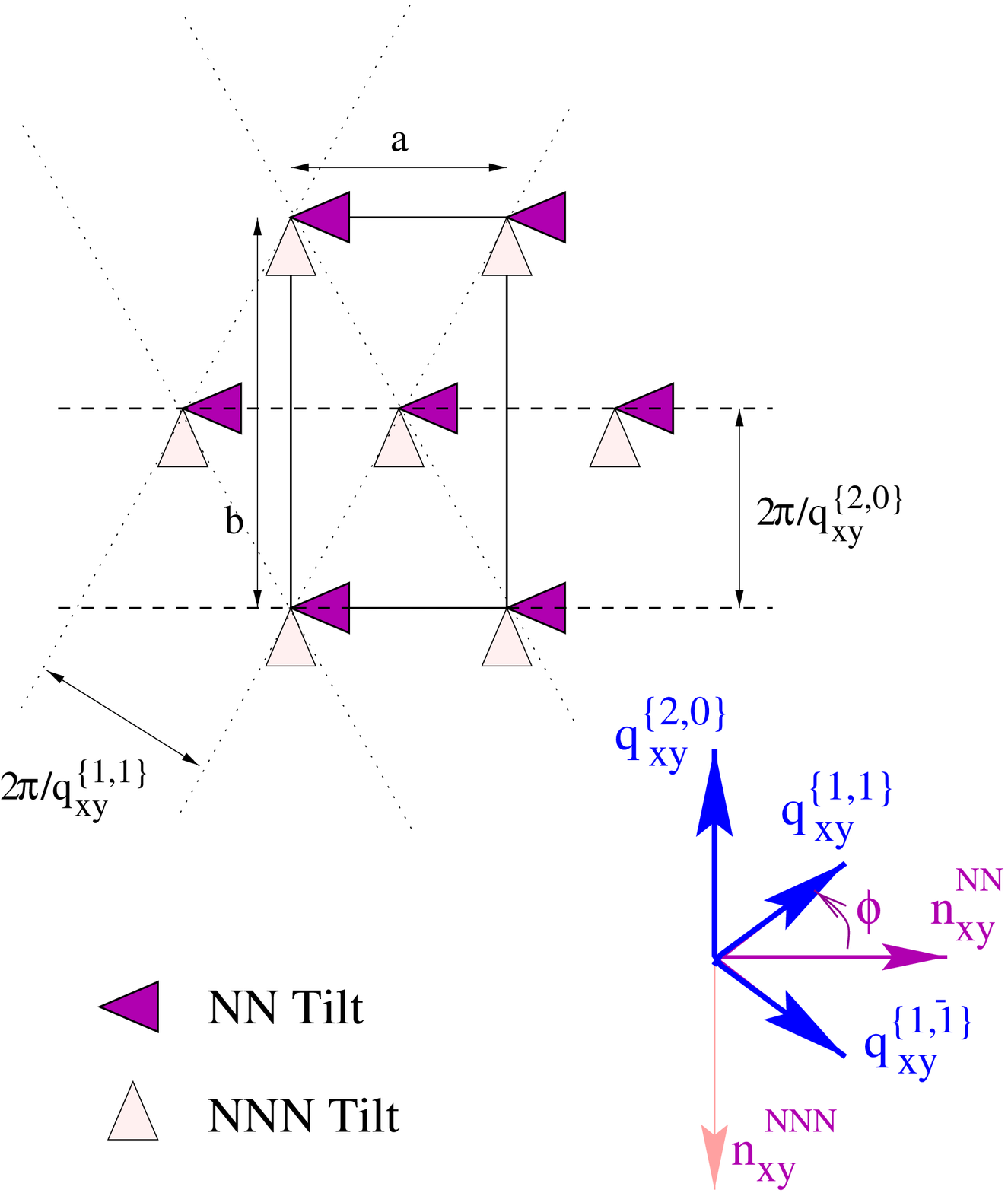} \hfil
\vskip 1truecm
\centerline{Fig. 4}
\newpage

\epsfxsize 12truecm
\hfil \epsfbox{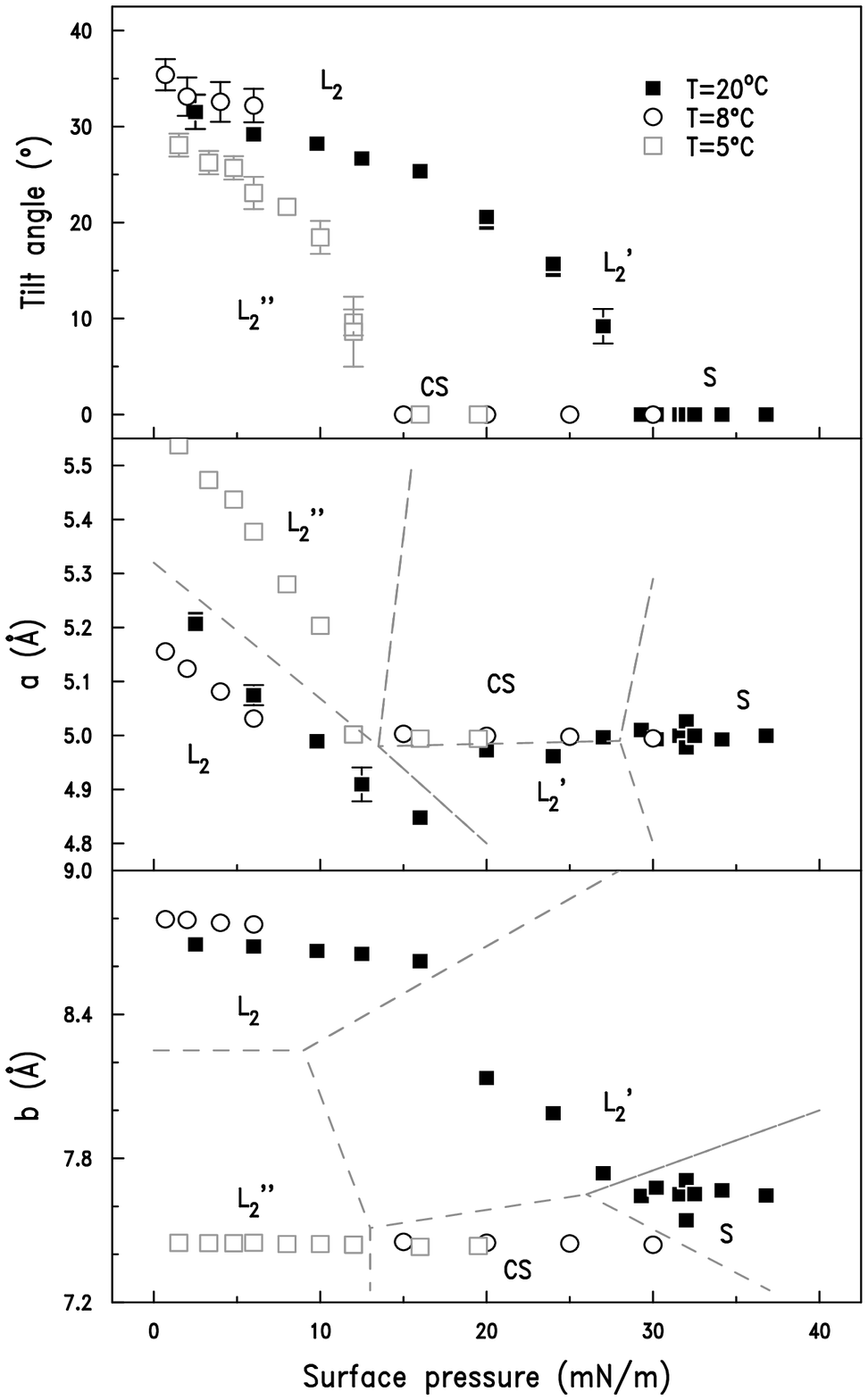} \hfil
\vskip 1truecm
\centerline{Fig. 5}
\newpage

\epsfxsize 12truecm
\hfil \epsfbox{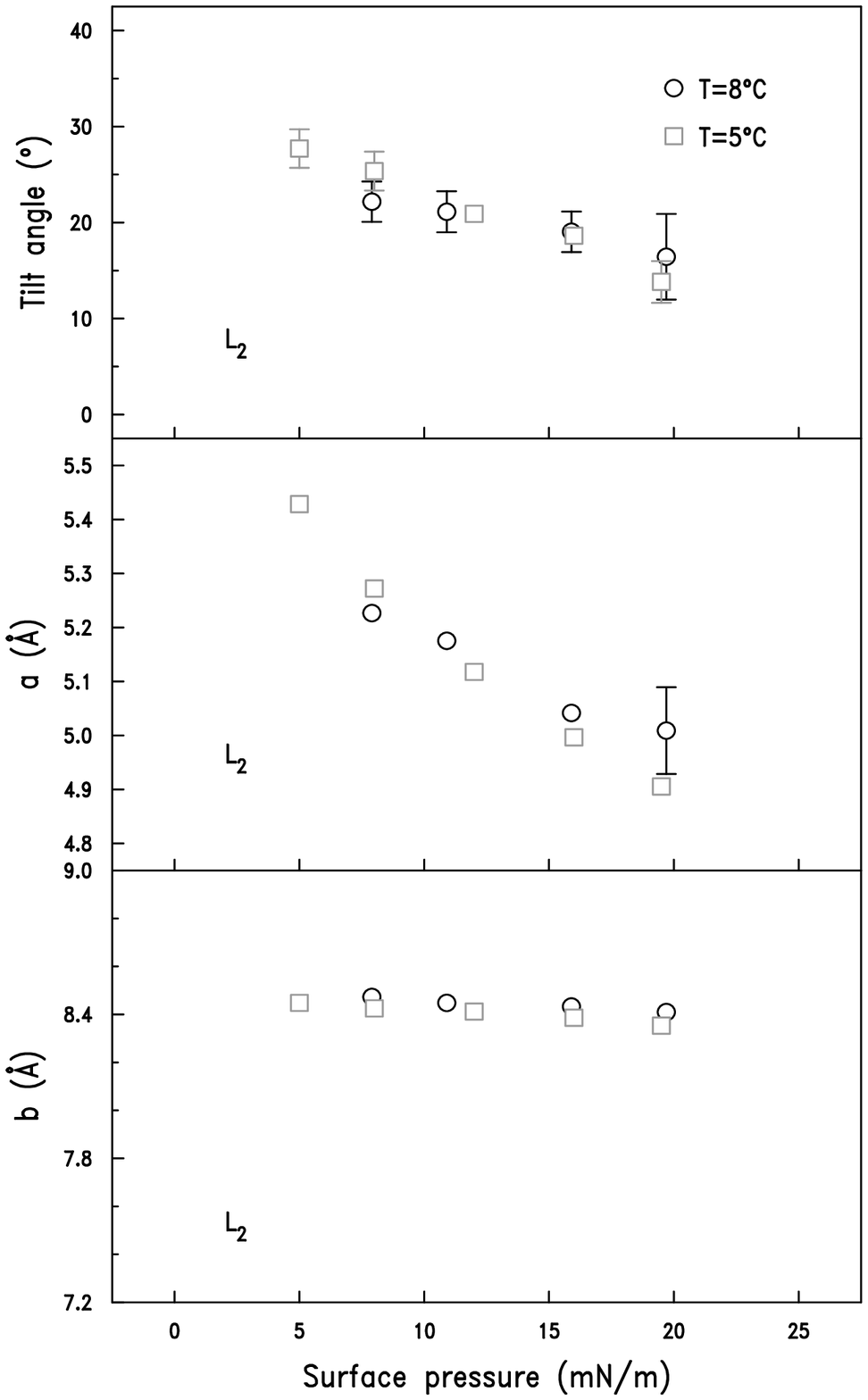} \hfil
\vskip 1truecm
\centerline{Fig. 6}
\newpage

\epsfxsize 14truecm
\hfil \epsfbox{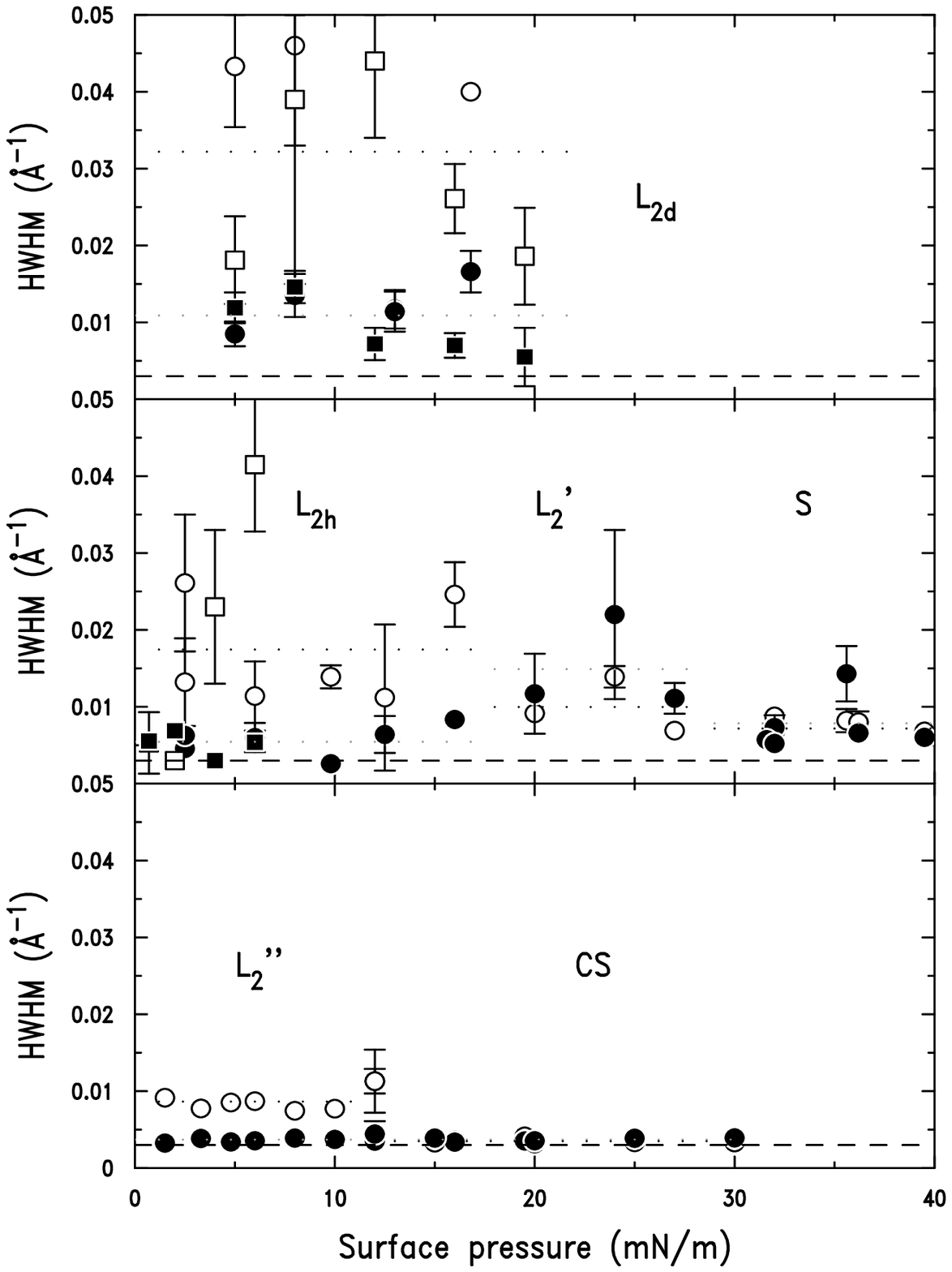} \hfil
\vskip 1truecm
\centerline{Fig. 7}
\newpage

\epsfxsize 14truecm
\hfil \epsfbox{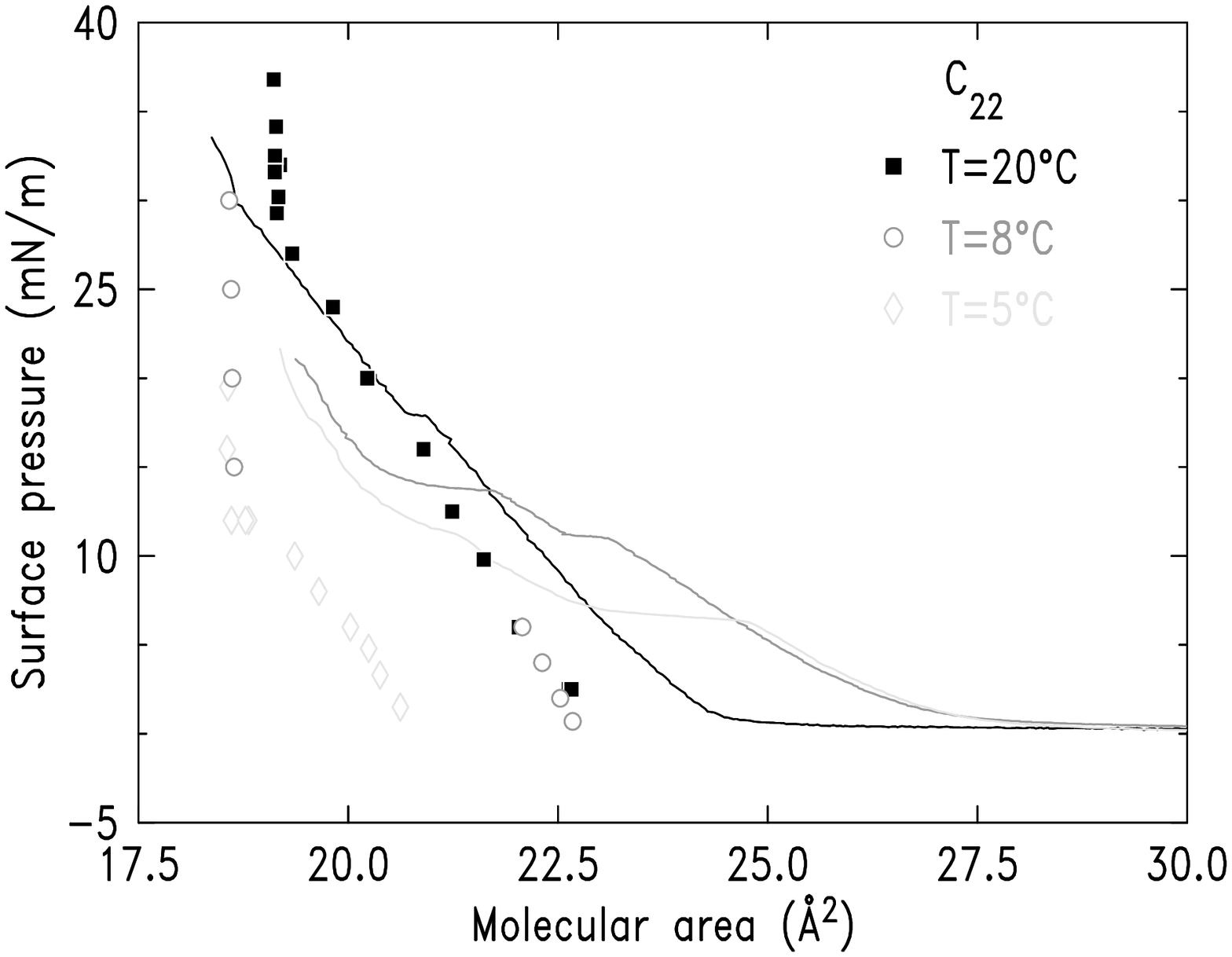} \hfil
\vskip 1truecm
\centerline{Fig. 8}
\newpage

\epsfxsize 14truecm
\hfil \epsfbox{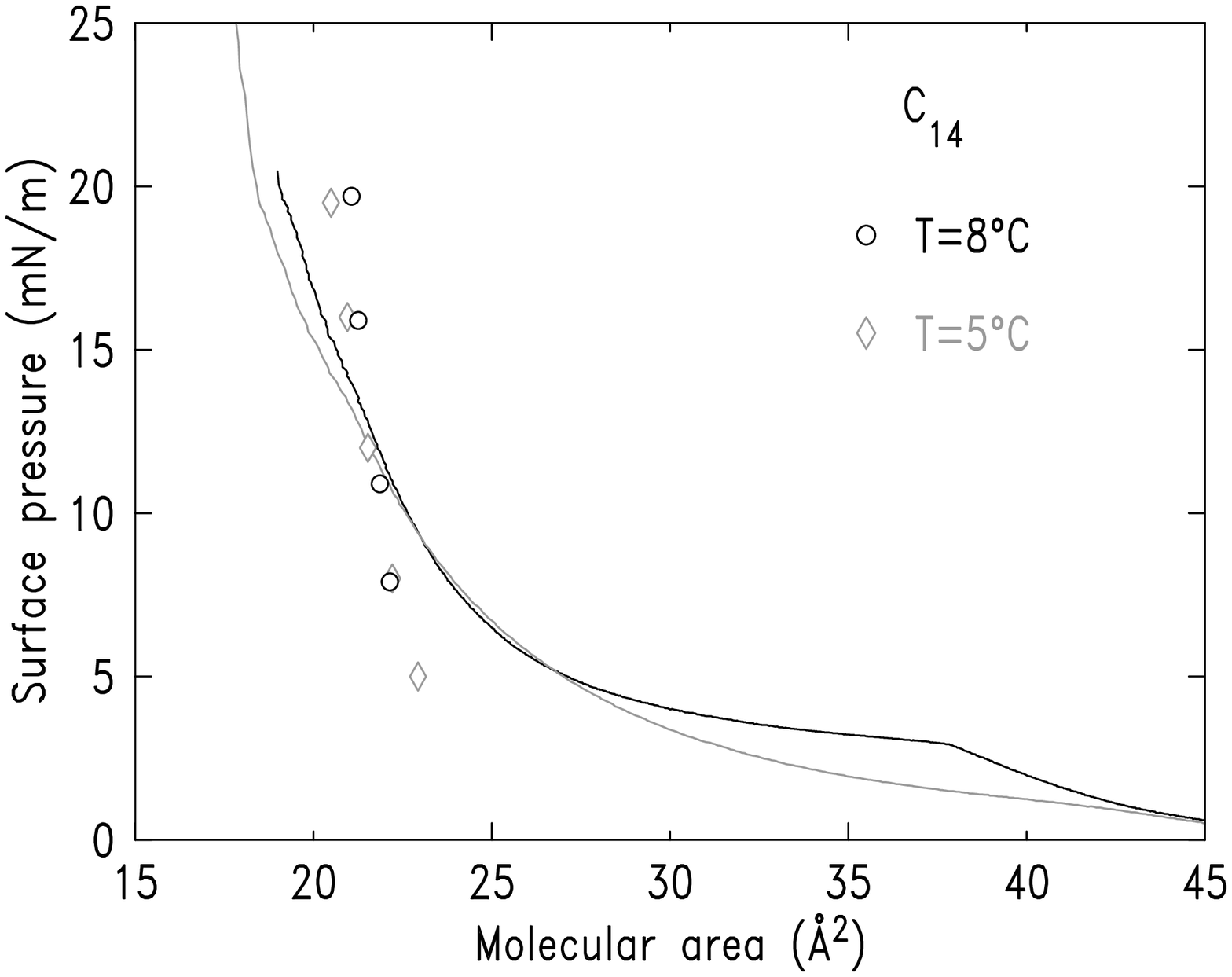} \hfil
\vskip 1truecm
\centerline{Fig. 9}
\newpage

\epsfxsize 14truecm
\hfil \epsfbox{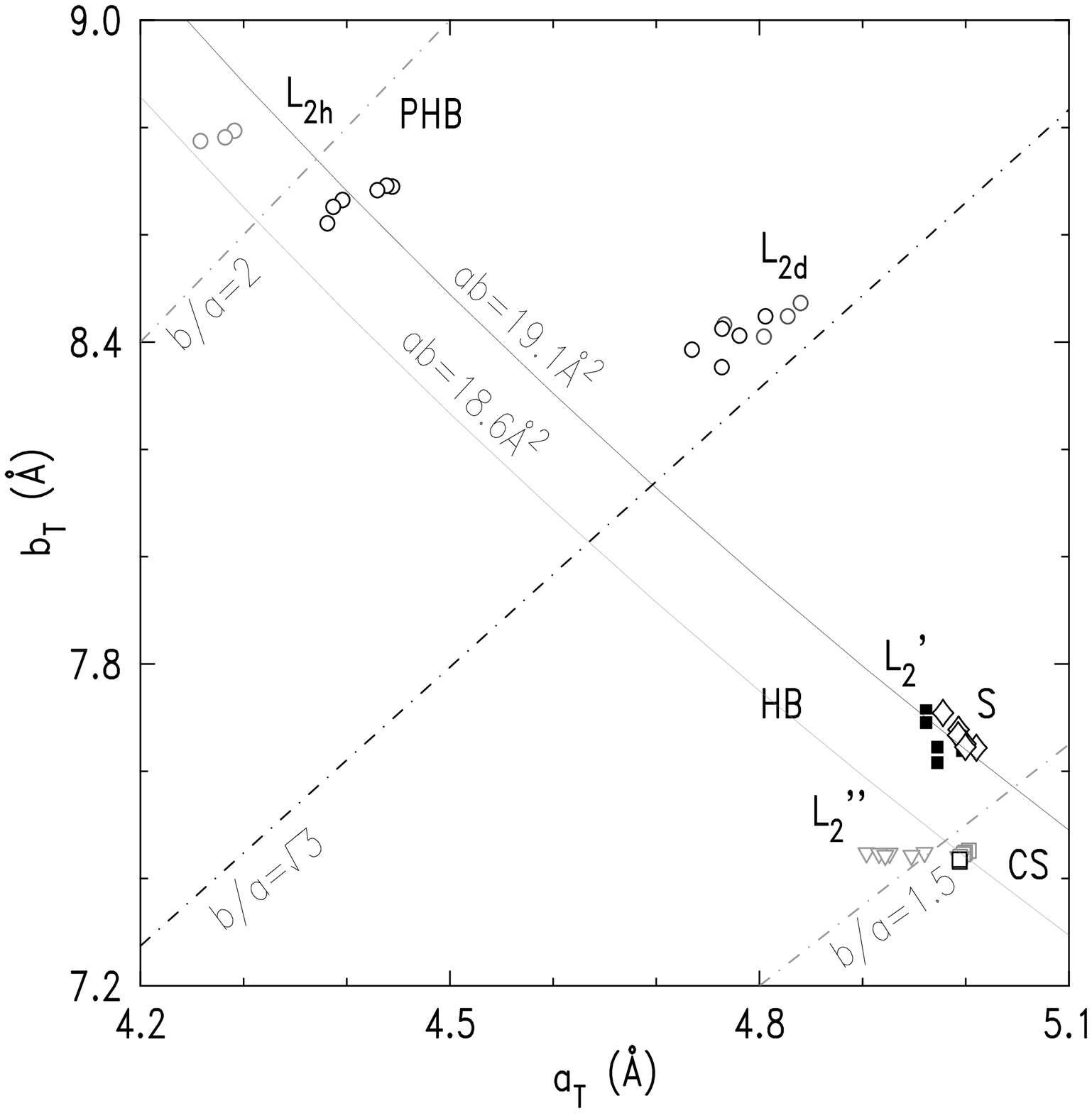} \hfil
\vskip 1truecm
\centerline{Fig. 10}
\newpage

\epsfxsize 14truecm
\hfil \epsfbox{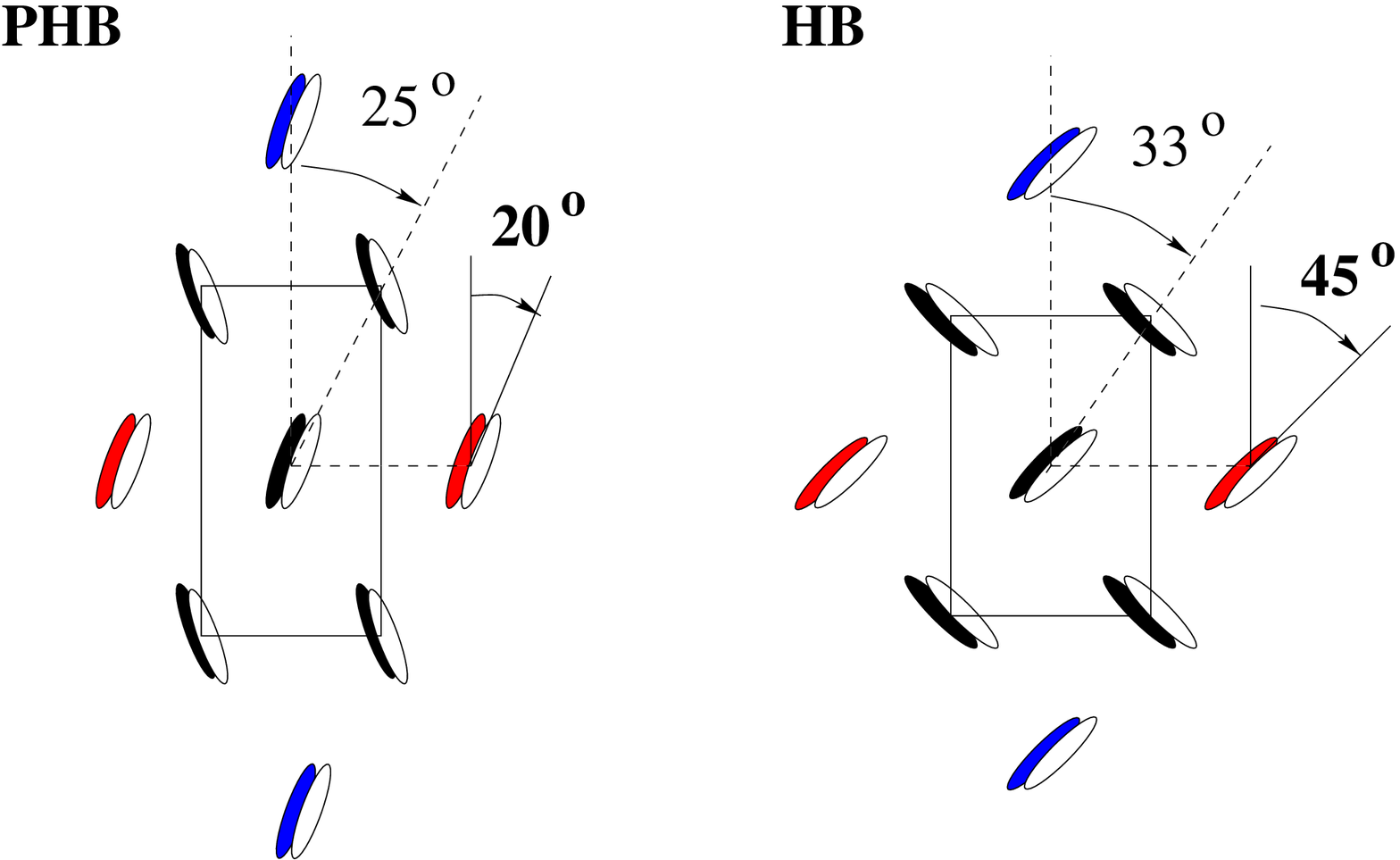} \hfil
\vskip 1truecm
\centerline{Fig. 11}
\newpage

\epsfxsize 14truecm
\hfil \epsfbox{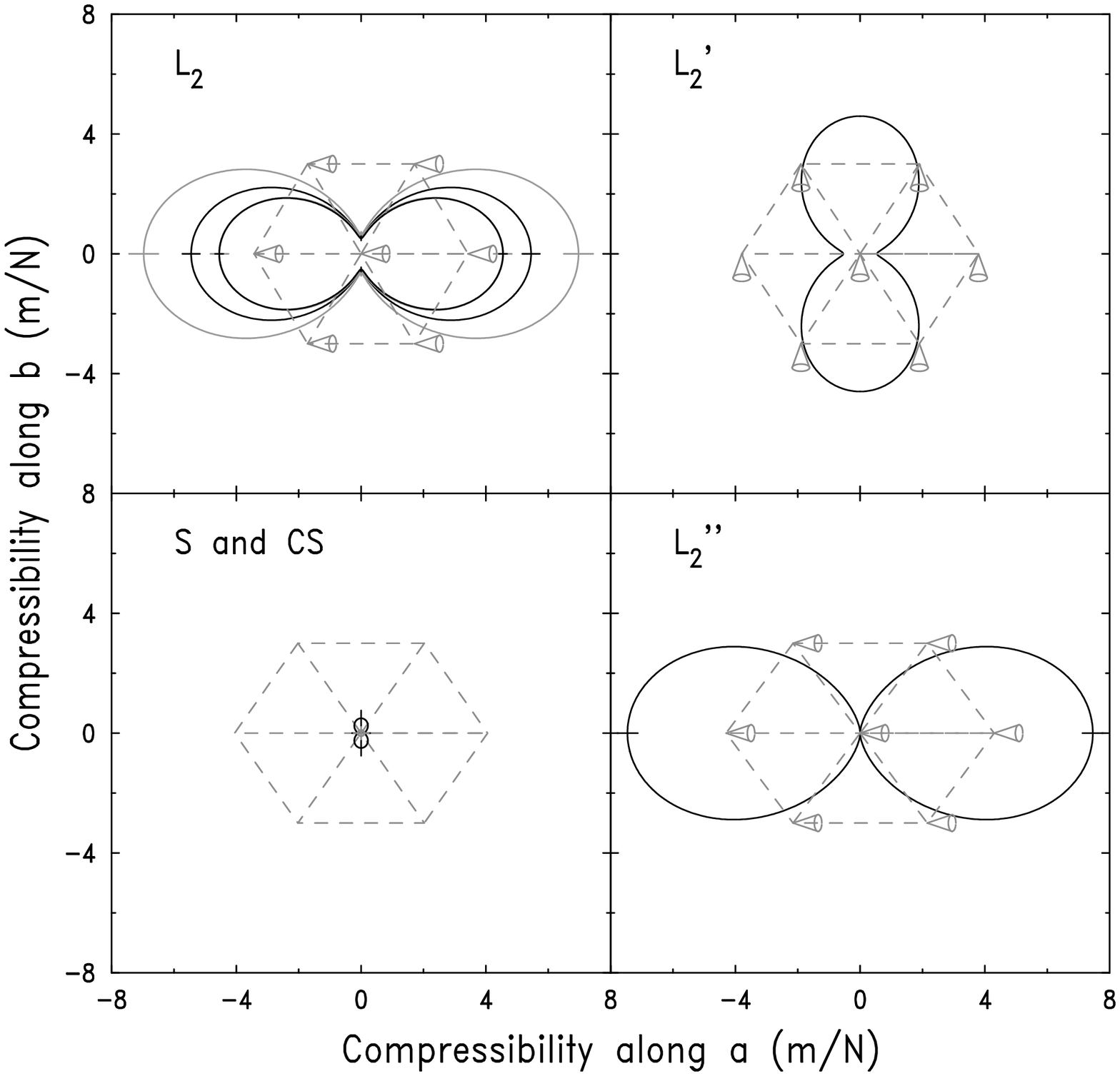} \hfil
\vskip 1truecm
\centerline{Fig. 12}
\newpage

\epsfxsize 14truecm
\hfil \epsfbox{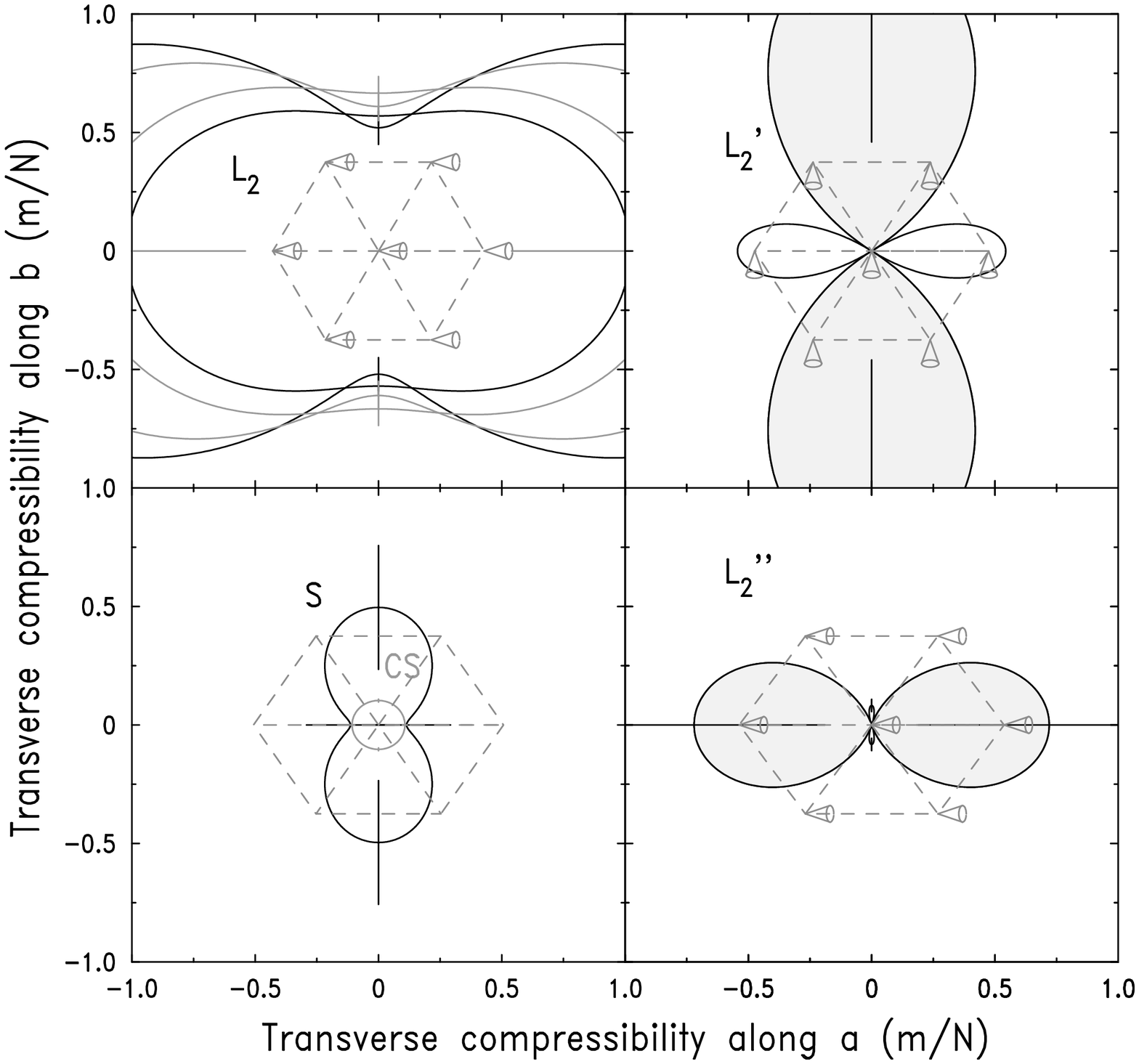} \hfil
\vskip 1truecm
\centerline{Fig. 13}
\newpage

\epsfxsize 14truecm
\hfil \epsfbox{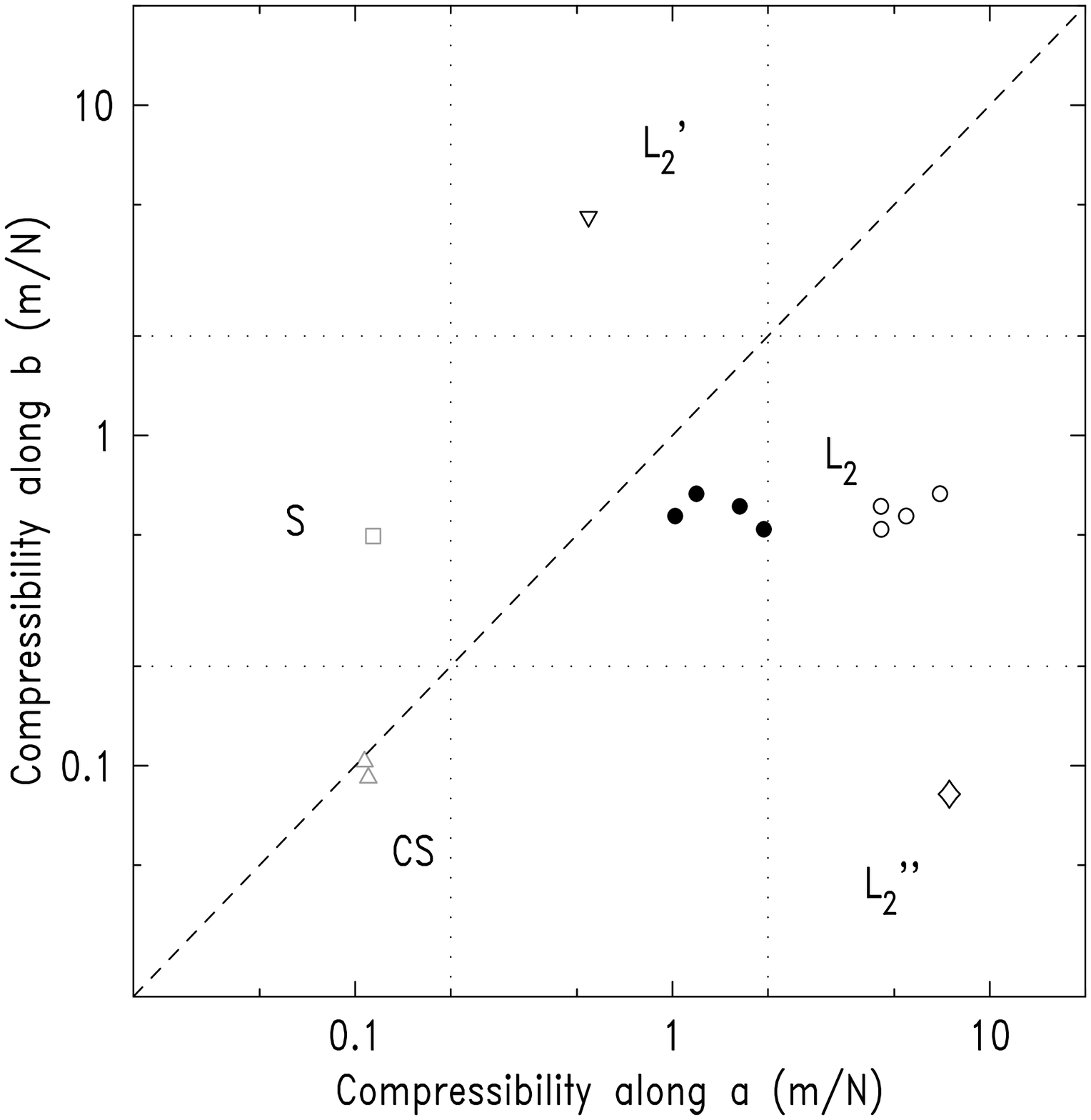} \hfil
\vskip 1truecm
\centerline{Fig. 14}
\newpage

\epsfxsize 14truecm
\hfil \epsfbox{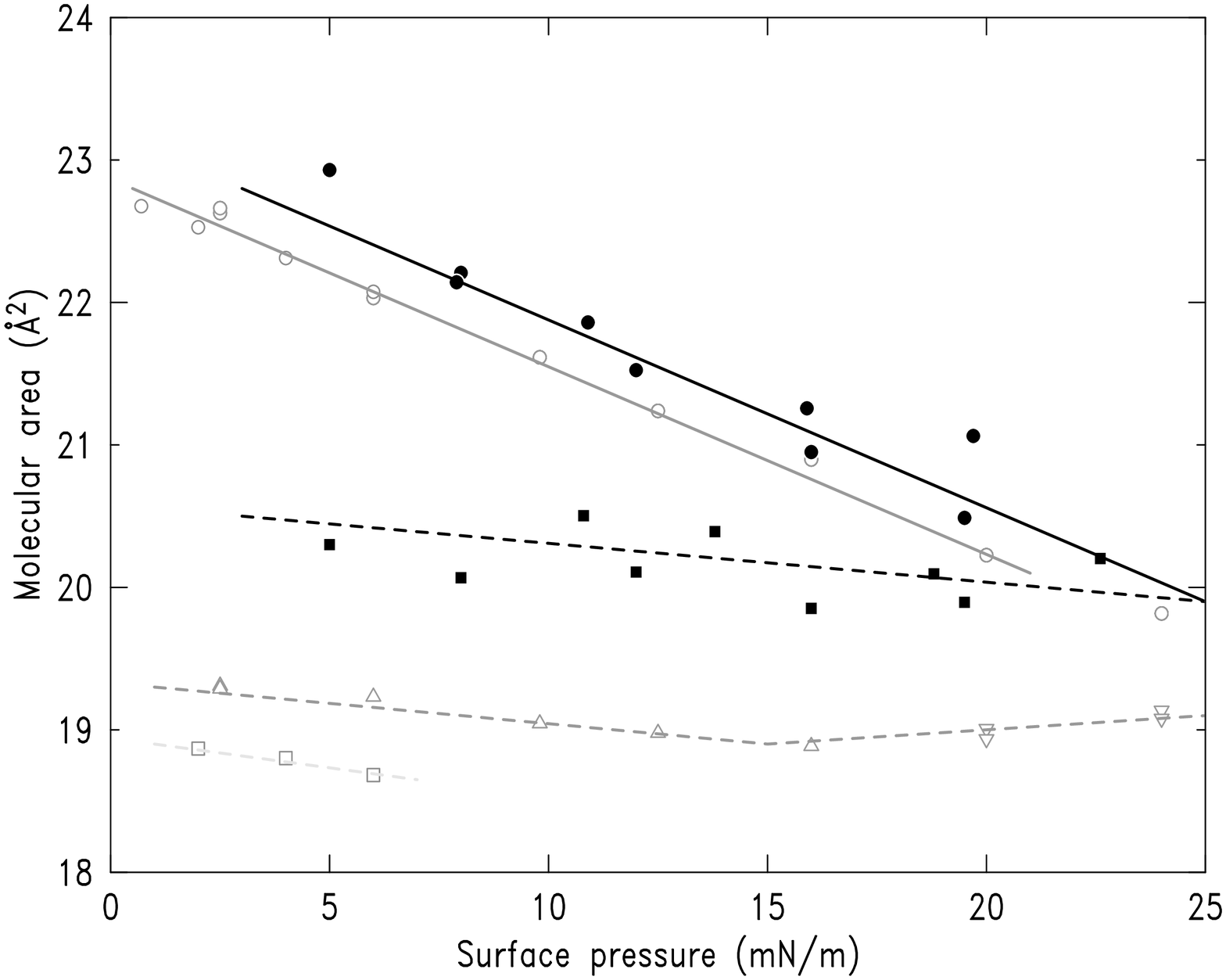} \hfil
\vskip 1truecm
\centerline{Fig. 15}
\newpage

\epsfxsize 14truecm
\hfil \epsfbox{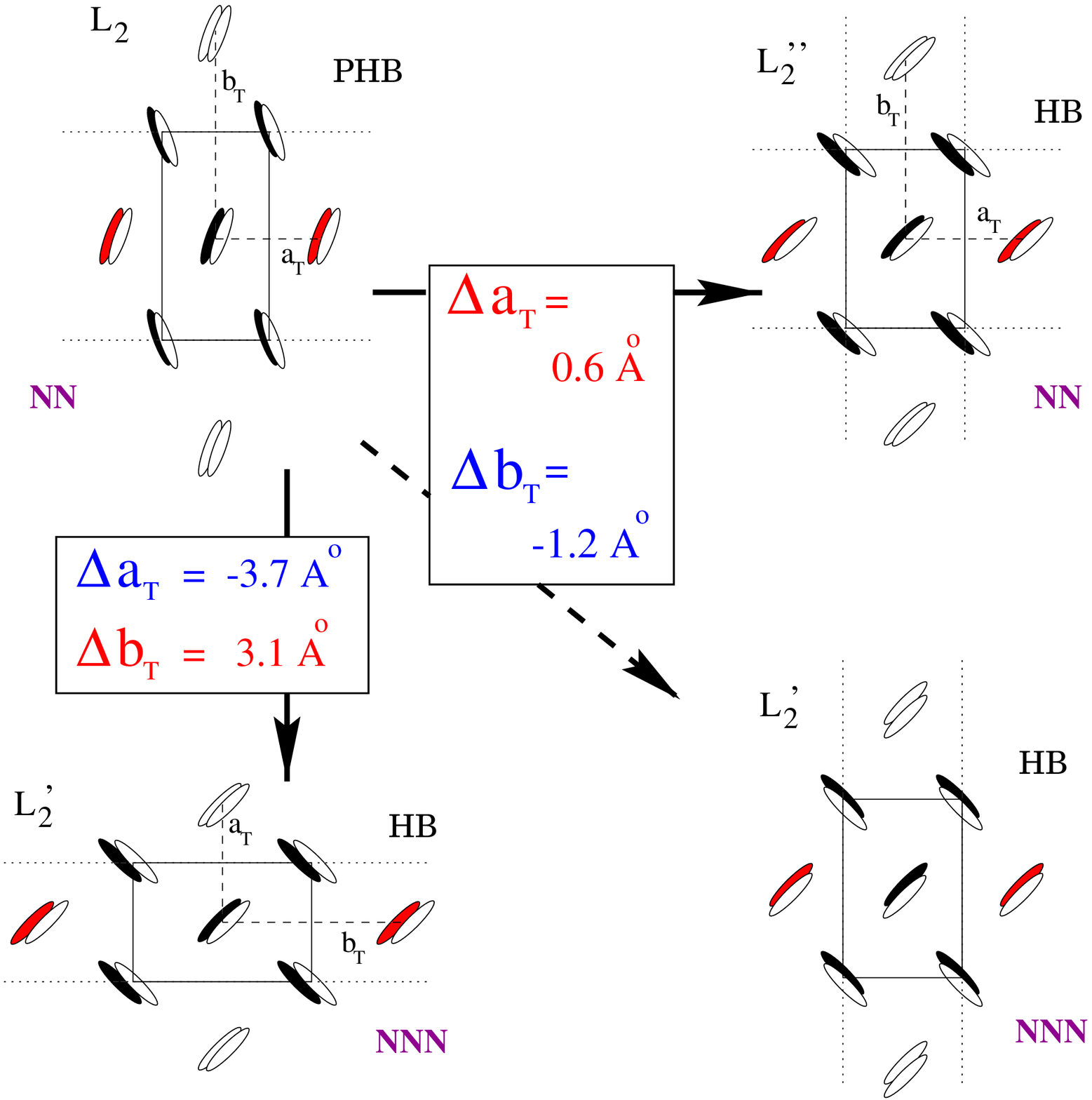} \hfil
\vskip 1truecm
\centerline{Fig. 16}
\newpage

\epsfxsize 14truecm
\hfil \epsfbox{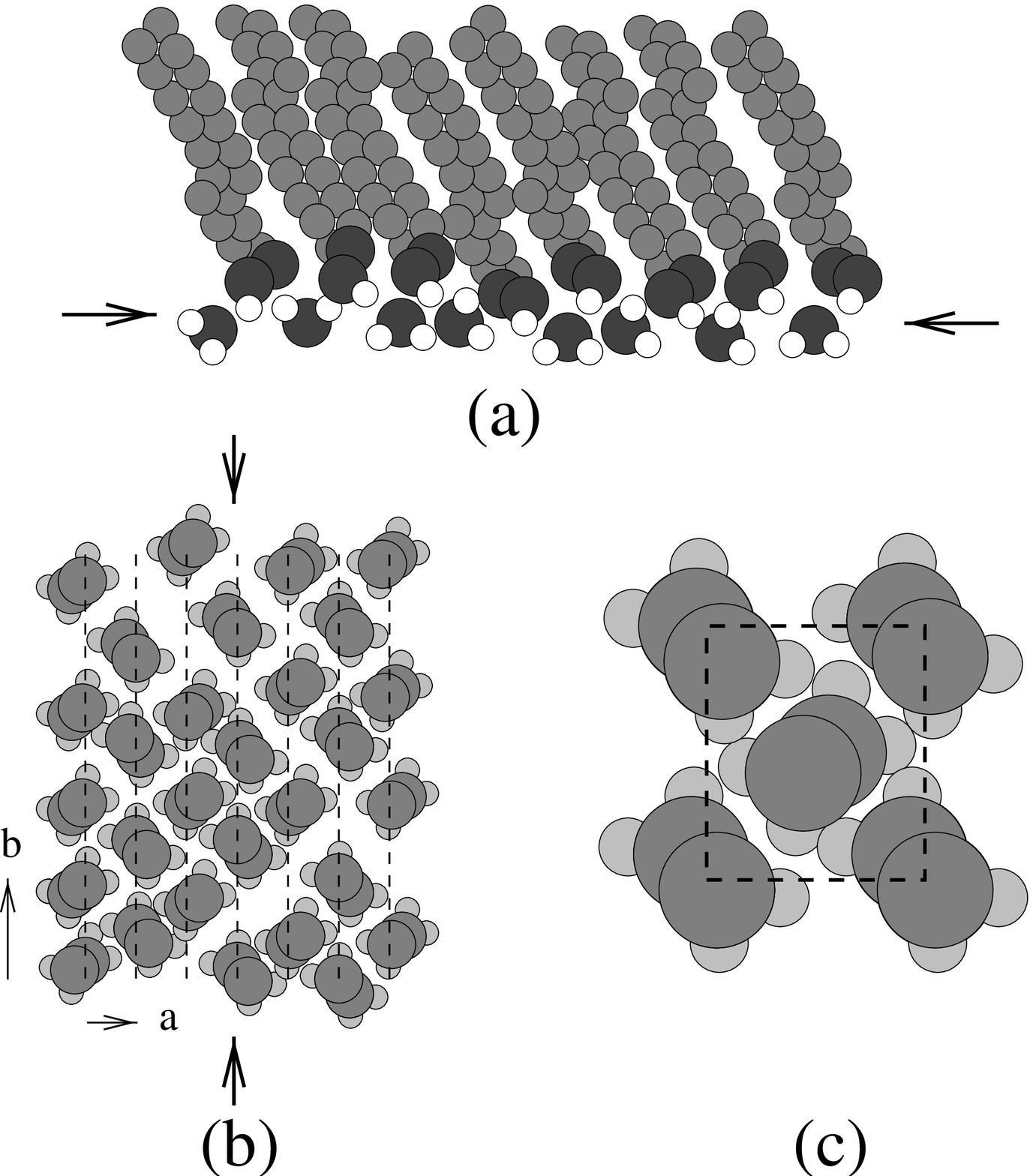} \hfil
\vskip 1truecm
\centerline{Fig. 17}
\newpage

\end{document}